\documentclass[twocolumn]{aastex63}
\usepackage{graphicx}
\usepackage{natbib}
\usepackage{placeins}
\usepackage{amsmath}
\usepackage{tgcursor}
\usepackage{booktabs}
\usepackage{enumitem}
\setcitestyle{notesep={ }}

\begin{document}

\title{Ice inheritance in dynamical disk models}

\author{Jennifer B. Bergner}
\affiliation{University of Chicago Department of the Geophysical Sciences, Chicago, IL 60637, USA}
\affiliation{NASA Sagan Fellow}

\author{Fred Ciesla}
\affiliation{University of Chicago Department of the Geophysical Sciences, Chicago, IL 60637, USA}

\begin{abstract}
\noindent The compositions of planet-forming disks are set by a combination of material inherited from the interstellar medium and material reprocessed during disk formation and evolution.  Indeed, comets and primitive meteorites exhibit interstellar-like isotopic ratios and/or volatile compositions, supporting that some pristine material was incorporated intact into icy planetesimals in the Solar Nebula.  To date, the survival of volatile interstellar material in the disk stage has not been modeled using realistic disk physics.  Here, we present a modeling framework to track the destruction of interstellar ices on dust grains undergoing transport processes within a disk, with a particular focus on explaining the incorporation of pristine material into icy planetesimals.  We find it is difficult to explain inheritance through the local assembly of comets, as ice destruction is rapid for small ($<$10$\mu$m) grains in the inner few tens of au.  Instead, a plausible pathway to inheritance is to form pebbles at larger disk radii, which then drift inwards to the comet-forming zone with their ices mostly preserved.  Small grains beyond $\sim$100 au can experience ice photodissociation at the tens of percent level, however little of the ice is actually lost from the grain, likely making this a robust site for in situ ice chemistry.  Our models also indicate that many complex organic species should survive passage through the disk intact.  This raises the possibility that organics synthesized in the interstellar medium can be delivered to terrestrial planets by icy body impact and thus potentially participate in origins of life chemistry.  

\end{abstract}


\section{Introduction}
\label{sec:intro}

The composition of a planet-forming disk is controlled by a combination of inheritance from the interstellar medium (ISM), and chemical reprocessing during the formation and evolution of the disk. 
Numerous lines of evidence support that at least some solar system material was inherited intact from the ISM.  For instance, the high D/H ratios measured in H$_2$O and H$_2$S for the comet 67P/Churyumov–Gerasimenko seem to necessitate a prestellar origin for some of the ices in the comet \citep{Altwegg2017, Rubin2020}.  Similarly, the organic compositions of comets match closely with the organic inventories of Sun-like protostars \citep{Bockelee-Morvan2000, Bergner2017, Drozdovskaya2019}, further suggesting an interstellar provenance for cometary ices.  The parent bodies of primitive meteorites, which formed at smaller disk radii and were exposed to enhanced processing compared to comets, have been estimated to preserve 5-10\% pristine interstellar material \citep{Alexander2017}.  Thus, current evidence suggests that significant reservoirs of interstellar ices survived the processes of disk formation and evolution to be incorporated intact into comets and, to a lesser extent, asteroids.  

Disentangling the contributions of pristine versus reprocessed material in disks is a fundamental step in understanding both the chemistry and physics associated with planet formation.  Chemically, the ISM is host to a rich assortment of volatile/organic molecules, many of which are also of prebiotic interest.  For instance, small molecules such as H$_2$O and HCN that are implicated in origins of life scenarios \citep[e.g.][]{Powner2009} are commonly detected along all stages of the star formation sequence \citep{Snyder1971, vanDishoeck1995, Dutrey1997, vanDishoeck2014}.  Moreover, larger molecules such as formamide (NH$_2$CHO) and glycolaldehyde (CH$_2$OHCHO) which share structural elements in common with present-day biomolecules have been detected in molecular clouds and protostars, though generally with low abundances \citep{Turner1989, Hollis2000, Jorgensen2012, Kahane2013}.  If prebiotically useful molecules synthesized in the ISM are incorporated into icy bodies in the disk, they may be delivered to terrestrial surfaces via impact, and could potentially play a role in jump-starting origins of life chemistry \citep{Oro1961, Anders1989, Alexander2012, Rubin2019}.  Indeed, modeling of D/H fractionation chemistry in the Solar Nebula suggests that tens of percent of the water in Earth's oceans may be interstellar in origin \citep{Cleeves2014}, supporting a scenario in which pristine interstellar molecules are efficiently delivered to planetary surfaces. 

Interstellar isotopic signatures also offer valuable clues into the physical conditions of the young Solar system.  For instance, compared to other elements, oxygen isotopic compositions are notably heterogenous in primitive meteorites.  The currently favored explanation for this is that CO self-shielding in the parent molecular cloud produced a reservoir of $^{17}$O- and $^{18}$O-rich H$_2$O ice \citep{Yurimoto2004, Krot2020}.  In the disk stage, differential dynamical evolution of the ice and gas phases, along with different degrees of thermal processing, resulted in a range of O fractionation levels incorporated into forming solids.  With this framework, the preservation vs. `reset' of prestellar oxgyen isotope signatures has been used to infer various physical properties of the protosolar disk \citep[e.g][]{Alexander2017} and the protosolar radiation environment \citep{Lee2008}. 

Numerous previous models have explored prestellar ice reprocessing during protostellar collapse and disk accretion \citep{Visser2009, Visser2011, Yang2013, Hincelin2013, Drozdovskaya2014, Drozdovskaya2016, Yoneda2016}.  These studies agree that significant reservoirs of prestellar ices are incorporated into the disk, especially for accretion trajectories at the outer edge of the disk.  While these simulations track the process of disk formation, the treatment of disk physics is simplified.  Importantly, dynamical processes such as dust settling, radial drift, and turbulence within the disk are not considered.  Based on observations of dust distributions in extrasolar disk systems \citep[e.g.][]{Testi2014} as well as the inferred transport of materials within our own Solar system \citep[e.g.][]{Bockelee-Morvan2002, McKeegan2006}, these dynamical effects are thought to profoundly reshape the disk physical structure during its evolution.  

Vertical transport within the disk should be particularly impactful on the chemistry, as disks exhibit steep vertical chemical gradients as a result of increased radiation shielding closer to the midplane \citep[e.g.][]{Aikawa1999}.  Indeed, particles undergoing turbulent mixing can be cycled between these vertical layers, resulting in exposure to very different thermal and UV environments \citep{Semenov2011, Ciesla2012}.  Properly treating disk dynamics is therefore critical to inferring interstellar ice survival within the disk environment.

In summary, there is compelling evidence that some interstellar material survived incorporation into icy Solar system bodies.  This has important implications for the potential delivery of prebiotic precursors to planetary surfaces, as well as for the physical conditions that characterized the Solar nebula.  Modeling of ice inheritance during protostellar infall and disk formation supports that significant reservoirs of pristine icy material should be incorporated into the disk.  However, to date the survival of interstellar ices within the disk stage has not been tested considering realistic disk physics.  In this work, we aim to explore the survival of pristine interstellar material within a dynamic disk environment.  To do so, we present simulations of icy grain trajectories subject to transport processes within the disk, and track the destruction of the ice mantles as particles are exposed to destructive thermal and photo-processing.  

Our modeling framework is outlined in Section \ref{sec:model}.  In Section \ref{sec:results} we present the resulting ice survival outcomes, considering the statistics for bulk ice survival as well as the preservation of specific molecules in the ice.  We discuss the implications for interstellar inheritance in Section \ref{sec:disc}, with a focus on the plausible incorporation if pristine ices into comets.  In Section \ref{sec:concl} we present our summary and conclusions.

\section{Model}
\label{sec:model}

\subsection{Disk structure}
\label{subsec:model_structure}
We adopt a parametric disk structure model based on the prescriptions of \citet{Andrews2011} and \citet{Rosenfeld2013} for an accretion disk \citep[see also][]{Lynden-Bell1974, Hartmann1998}.  We implement a disk structure informed by the well-characterized disk around the nearby T Tauri star TW Hya.  This allows us to simulate a Solar Nebula analog (M$_\star$ = 0.8M$\odot$) with well-constrained physical properties, whereas the structure of the Solar Nebula itself is not known in detail.  We assume the stellar properties for TW Hya of $T_\mathrm{eff}$=4110 K, $M_\star$=0.8 $M_\odot$, and $R_\star$ = 1.04 $R_\odot$ \citep{Andrews2012}.  The dust distribution is described as follows and is based on the best-fit dust model reported in \citet{Cleeves2015}.  A summary figure of the disk physical model is shown in Appendix \ref{sec:app_structure}.

The dust surface density profile $\Sigma_{\mathrm{dust}}$ is found from the tapered power-law:
\begin{equation}
    \Sigma_\mathrm{dust} (r) = 0.04\mathrm{\;g\;cm^{-2}} \Big{(}\frac{r}{150 \mathrm{au}}\Big{)}^{-1} \mathrm{exp} \Big{(}-\frac{r}{150 \mathrm{au}}\Big{)}.
\end{equation}
We include two dust populations to capture the effects of dust growth and settling within the disk.  The small-grain population ranges in size from  5 nm--10 $\mu$m, and the large-grain population from 5 nm--1 cm.  We assume 90\% of the total dust mass is contained in the large grain population, or $X_{\mathrm{lg}}=0.9$.  Both grain populations extend from an inner radius of 0.05 au to an outer radius of 200 au.  The volumetric densities of each grain population $i$ follow a Gaussian distribution with disk height $z$:
\begin{equation}
    \rho_{\mathrm{dust},i} (r,z) = X_i\frac{\Sigma_{\mathrm{dust}}(r)}{\sqrt{2\pi}H_i(r)} \mathrm{exp}\Big{[}-0.5\Big{(}\frac{z}{H_i(r)}\Big{)}^2\Big{]},
\end{equation}
where $H_i$ is the vertical scale height of each dust population.  The small grain scale height is described by a power-law:
\begin{equation}
H_{\mathrm{sm}}(r) = 15 \mathrm{\;au} \Big{(}\frac{r}{150\mathrm{au}}\Big{)}^{1.3} .
\end{equation}
To mimic settling, the large grain scale height $H_{\mathrm{lg}}(r)$ is set to 0.2$H_{\mathrm{sm}}(r)$.

We solve for the local dust temperatures in the disk using the radiative transfer code RADMC-3D \citep{Dullemond2012} including isotropic scattering.  Dust opacities are calculated using the DIANA project \texttt{Opacity-Tool} \citep{Woitke2016} assuming an amorphous silicate-carbonaceous material with a carbon fraction of 0.13 and a size power-law distribution of 3.5.  The input stellar spectrum is composed of the TW Hya UV spectrum \citep{Herczeg2002, Herczeg2004, Cleeves2013} added to a Blackbody component calculated from the stellar $T_\mathrm{eff}$=4110 K.  We additionally include an external UV field of 30 $G_0$, where $G_0$=1 corresponds to the ISRF UV (91--200 nm) flux of 1.6$\times$10$^{-3}$ erg cm$^{-2}$ s$^{-1}$ \citep{Habing1968}.  In this model the external UV field only becomes important beyond $\sim$180 au, and therefore does not play a significant role in ice destruction at the radii we are considering ($<$150 au).  For disks with a stronger external radiation field, ice loss may be enhanced towards the outer disk edge compared to our models.

We also use RADMC-3D to sample the local UV field at each grid point in our disk model.  Fluxes are computed at 71 wavelengths spanning 91-200 nm.  The wavelengths are chosen such that, given the unattenuated stellar spectrum, the total wavelength-summed photodissociation rate differs by less than 5\% from the wavelength-integrated photodissociation rate for all molecules considered in our model (see Section \ref{sssec:photodissociation}).  Note that our treatment does not include resonant scattering effects on the propagation of Ly-$\alpha$ photons, which has been shown to enhance Ly-$\alpha$ penetration in deeper disk layers \citep{Bethell2011}.  As such, the rates of photodesorption and photodissociation may be somewhat underestimated in our treatment.

Gas temperatures are calculated using the RADMC-3D derived dust temperatures.  We follow the formalisms of \citet{Dartois2003} and \citet{Rosenfeld2013}, using the parametric values provided for TW Hya in \citet{Zhang2017} and \citet{Huang2018}.  The gas temperature at the midplane is set equal to the dust temperature:
\begin{equation}
    T_{\mathrm{gas}}(r, z=0) = T_{\mathrm{dust}}(r, z=0) \equiv T_m
\end{equation}
Above a disk elevation $z_q$, the gas temperature is:
\begin{equation}
    T_{\mathrm{gas}}(r, z\geq z_q) = 125\mathrm{K} \Big{(}\frac{r}{10 \mathrm{au}}\Big{)}^{-0.47} \equiv T_a
\end{equation}
At $0<z<z_q$, the gas temperature is described by:
\begin{equation}
    T_{\mathrm{gas}}(r,0<z<z_q) = T_a + (T_m - T_a) \Big{(}\mathrm{cos}\frac{\pi z}{2 z_q} \Big{)}^{2\delta},
\end{equation}
where $\delta$ is the shape of the vertical gas temperature gradient and is set to 2.0.  $z_q$ is set to 4$\times$ the gas scale height $H_{\mathrm{gas}}$, where:
\begin{equation}
    H_{\mathrm{gas}} = \Big{(}\frac{kT_mr^3}{GM_\star \mu m_H}\Big{)}^{1/2}.
\end{equation}
$\mu$ is the mean molecular gas weight, set to 2.37.  We then calculate the gas density by solving the hydrostatic equation given a vertically integrated gas to dust ratio of 100:1.

\begin{figure*}
    \includegraphics[width=\linewidth]{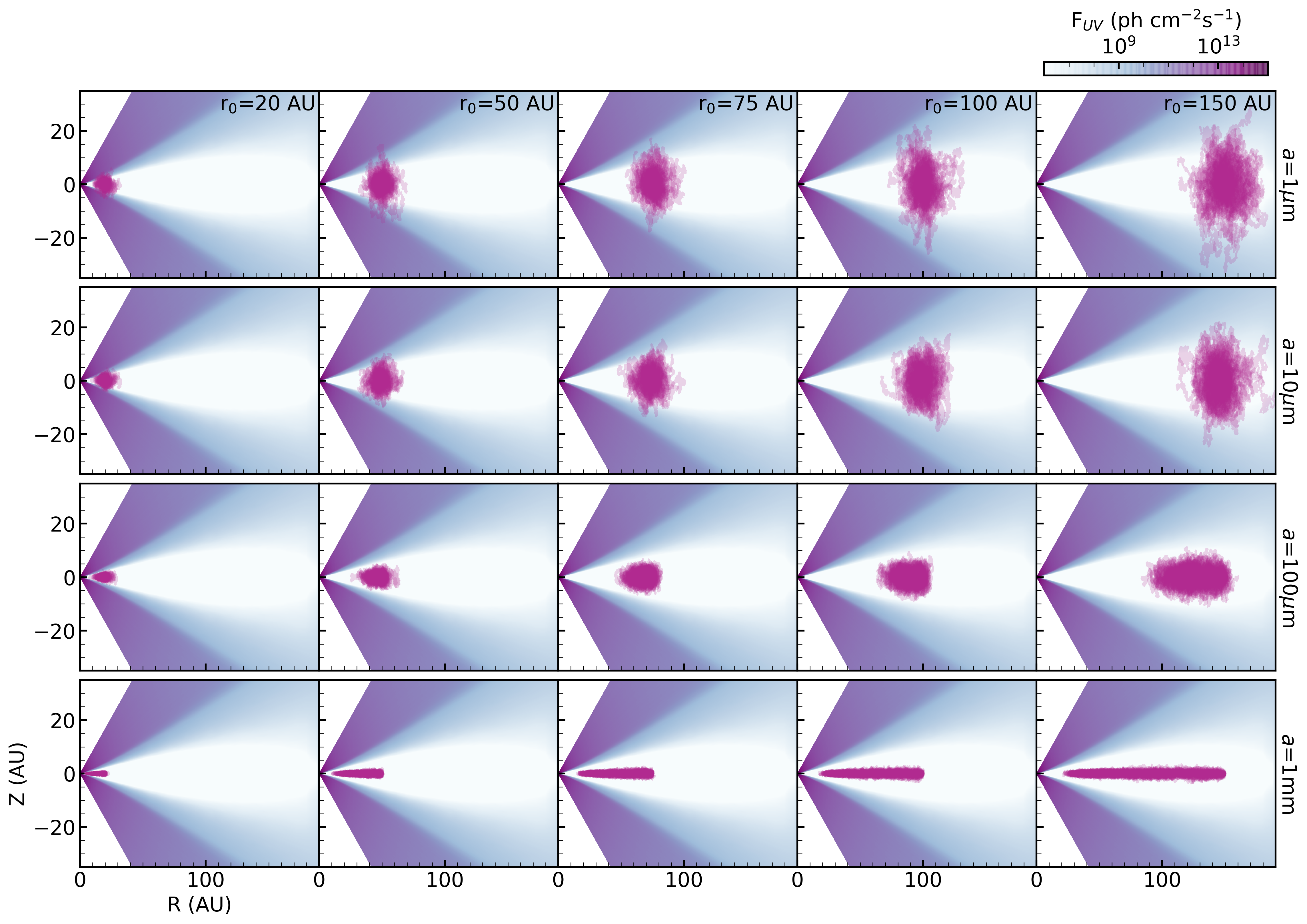}
    \caption{`Clouds' (pink) showing 100 particle trajectories for different sized particles initialized at different radii and run for 100 kyr. Trajectory clouds are overlaid on the modeled UV field.}
    \label{fig:traj_cloud}
\end{figure*}

\subsection{Particle dynamics}
\label{subsec:dynamics}

We simulate the dynamical evolution of solid grains within the disk using particle tracking methods \citep{Ciesla2010,Ciesla2011,Ciesla2012}.  The disk is assumed to be turbulent, imparting a local diffusivity within the gas given by $\alpha c_{s} H$.  For the models presented here, we take $\alpha$=10$^{-3}$ everywhere in the disk.

We consider particles with radii of 1, 10, 100, and 1000 $\mu$m, representing relatively unevolved grains through to grains that have grown significantly within the disk.  In each case we adopt a density of 1.5 g cm$^{-3}$.  The particles are released at 20, 50, 75, 100, and 150 au, with their dynamical evolution determined by the combination of turbulent diffusion, gravitational settling, and gas drag.  The movement of each particle is calculated as described in \citet{Ciesla2010} and \citet{Ciesla2011}, and positions ($r$,$z$) recorded every timestep where $\Delta t$ is taken to be 1/25$^{\mathrm{th}}$ the local orbital period.  

Particles are initialized at the disk midplane.  This is motivated by the results of protostellar infall models \citep[e.g.][]{Visser2009, Hincelin2013}, which show that significant reservoirs of intact interstellar ices are incorporated in the disk as it forms due to a combination of (i) shallow accretion trajectories and preferential accretion at the disk edge, and (ii) the presence of the envelope which shields infalling material from photoprocessing.  Thus, we assume that pristine interstellar ice has already survived the formation of the disk and is present on midplane grains.

Figure \ref{fig:traj_cloud} shows 100 particle trajectories for each combination of grain size and initial radius considered in our models, where each trajectory has been simulated for 100 kyr.  The trajectory `clouds' reveal the typical dynamical behavior of different particle sizes.   The modeled UV field is also shown to illustrate the disk regions where particles are subject to ice processing.  Small particles (1 and 10$\mu$m) are typically mixed into elevated disk regions where they are exposed to moderate to high UV fluxes, and remain roughly centered around their initial radius.  100$\mu$m particles experience some vertical mixing but are still mostly confined to UV-shielded disk layers.  These particles also experience slow drift inwards towards the star.  1mm grains do not diffuse out of the disk midplane, and also experience more rapid drift towards the star.  These differences in dynamical evolution will be largely responsible for differences in ice survival outcomes in our subsequent analysis.

\subsection{Chemical model}
\label{subsec:model_chem}
Our chemical modeling framework simulates a protostellar-like icy grain structure and composition and treats the destructive processes of thermal desorption, photodesorption, and photodissociation.  In this work, we focus solely on ice destruction with the specific aim of identifying the degree to which protostellar ice can survive passage through the disk.

\begin{figure*}
    \includegraphics[width=\linewidth]{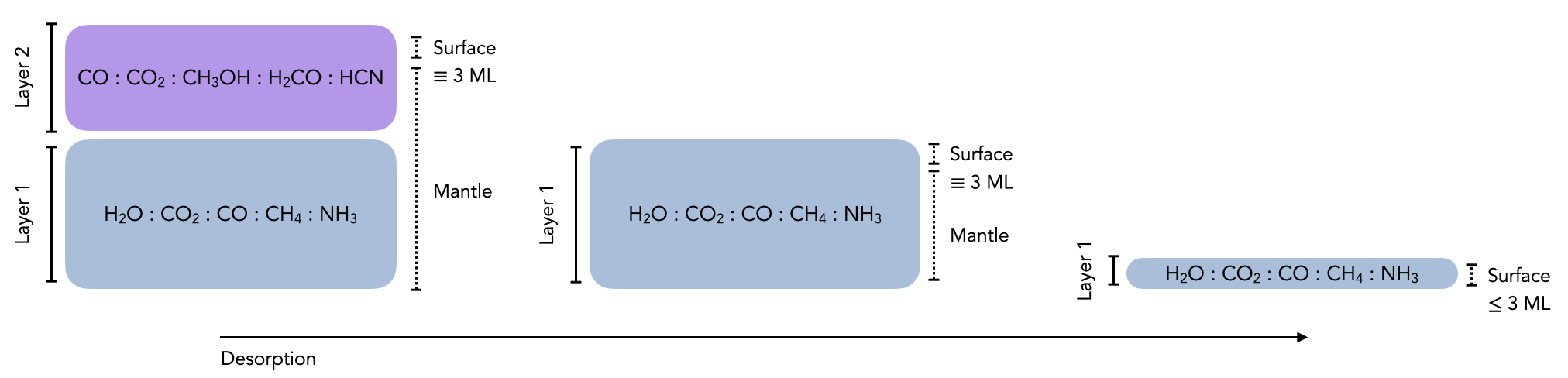}
    \caption{Schematic of ice model structure and its evolution as ice is lost due to desorption.  The bottom layer (Layer 1) represents first-generation protostellar ice species and the top layer (Layer 2) represents second-generation ice species.  Desorption takes place from the top 3 ML of the ice, referred to as the surface.  The non-surface is referred to as the mantle.  Once Layer 2 is lost, the surface zone moves into Layer 1.}
    \label{fig:ice_schematic}
\end{figure*}

The ice model, shown schematically in Figure \ref{fig:ice_schematic}, consists of two vertical layers, mimicking the ice compositions inferred along protostellar lines of sight \citep{Oberg2011}.  The total ice thickness is 100 monolayers (ML; 1 ML = 10$^{15}$ molecules cm$^{-2}$).  The bottom layer is a H$_2$O-dominated mixture with a composition 58 H$_2$O : 11.6 CO : 7.5 CO$_2$ : 2.9 CH$_4$ : 2.9 NH$_3$.  This represents `first-generation' ice species which form in the early stages of a molecular cloud.  The top layer is a CO-dominated mixture with a composition of 5.2 CO : 9.3 CO$_2$ : 1.7 CH$_3$OH : 0.2 HCN : 0.6 H$_2$CO.  This represents the `second-generation' ice species which form once cloud densities are high enough for catastrophic CO freeze-out.  The following sections outline our treatment of destructive ice processing, which draws from the models of \citet{Fayolle2011}, \citet{Ruaud2019}, and \citet{Garrod2019}.

\subsubsection{Thermal desorption}
\label{sssec:tdes}
The rate of molecule loss from the ice surface due to thermal desorption is found from
\begin{equation}
    R_{\mathrm{tdes},i} = -N_{i,s}\; \nu_{0,i}\; \;\mathrm{exp} (-E_{\mathrm{des},i}/T),
\end{equation}
where for species $i$, $N_{i,s}$ is the surface column density, $E_{\mathrm{des},i}$ is the binding energy, T is the local temperature, and $\nu_{0,i}$ is the pre-exponential factor akin to an escape frequency.  In the disk regions relevant to this study the temperatures of the small and large grain populations (Section \ref{subsec:model_structure}) differ by at most a few K, and we adopt the small grain temperature for calculating thermal desorption rates. For small molecules, $\nu_{0,i}$ can be found from the harmonic oscillator relation \citep[e.g.][]{Hasegawa1992}:
\begin{equation}
    \nu_{0,i} = \sqrt{\frac{2 N_s k_B E_{\mathrm{des},i}}{\pi^2 m_i}},
\end{equation}
where $N_s$ is the binding site density (10$^{15}$ cm$^{-2}$), $k_B$ is the Boltzmann constant, and $m_i$ is the mass of species $i$.  For the molecules considered here, $\nu_0$ is of order a few $\times$10$^{12}$ s$^{-1}$.

Our adopted binding energies are listed in Table \ref{tab:E_bind}, and are taken from existing chemical modeling frameworks \citep{Garrod2013, McElroy2013}.  We note that for NH$_3$, HCN, and H$_2$CO there are $>$50\% differences between these binding energies and those measured in laboratory experiments \citep{Sandford1993, Noble2012, Noble2013}.  Still, we use the chemical model values to ensure uniformity between the relative binding energies of different molecules and radicals in our model, particularly because of a lack of laboratory measurements for radical binding energies.

\begin{deluxetable}{lclc}
	\tabletypesize{\footnotesize}
	\tablecaption{Adopted binding energies \label{tab:E_bind}}
	\tablecolumns{4} 
	\tablewidth{\textwidth} 
	\tablehead{
		\colhead{Molecule}       & 
		\colhead{$E_\mathrm{des,i}$ (K)} &
        \colhead{Radical fragments}       & 
		\colhead{$E_\mathrm{des,i}$ (K)}}
\startdata
H$_2$O & 5700 & OH-H & 2850 \\
CO$_2$ & 2575 & CO-O & 1150\\
CO & 1150 & C-O & 800 \\ 
CH$_4$ & 1300 & CH$_3$-H & 1175 \\
NH$_3$ & 5530 & NH$_2$-H & 3960 \\
CH$_3$OH & 5530 & CH$_3$-OH & 2850\\ 
HCN & 2050 & CN-H & 1600\\ 
H$_2$CO & 2050 & HCO-H & 1600 \\
\enddata
\tablenotetext{}{Radical products are assumed to desorb with the binding energy of the heavier species.  When available, binding energies are taken from \citet{Garrod2013}.  In other cases (O, HCN, and CN) we adopt the UMIST recommended binding energies \citep{McElroy2013}.}
\end{deluxetable}

\subsubsection{Photodesorption}
We also consider loss of ice from the surface due to photodesorption.  We assume a constant photodesorption yield $Y_{\mathrm{pdes},i}$ of 10$^{-4}$ photon$^{-1}$ for all ice species, informed by the simulations of \citet{Andersson2008}.  The choice of a uniform $Y_{\mathrm{pdes}}$ follows the reasoning outlined in \citet{Ruaud2016}, particularly that photodesorption is often driven by photon absorption in the ice layer beneath the desorbing molecule, and it is therefore difficult to assign a molecule-specific photodesorption yield \citep[e.g.][]{Munoz-Caro2010, Bertin2012}.  Given  a standard geometric cross-section $\sigma_{\mathrm{pdes}}$ of 10$^{-15}$ cm$^{-2}$, the photodesorption rate can then be written as
\begin{equation}
    R_{\mathrm{pdes},i} = -N_{i,s}\; Y_{\mathrm{pdes},i}\; \sigma_{\mathrm{pdes}}\; F_{UV}/2.
\end{equation}
where $F_{UV}$ is the total 91--200 nm flux (Section \ref{subsec:model_structure}) and the factor 1/2 reflects that the surface is only exposed to photons from one direction.

\subsubsection{Replenishment of the desorption zone}
In practice, we assume that desorption takes place only from the top 3 ML of the ice.  When molecules are lost from this surface layer through either thermal or photo-desorption, the empty positions are replenished statistically based on the composition of the layer underneath.  In the initial stage of ice loss, the surface is replenished from the Layer 2 mantle (Figure \ref{fig:ice_schematic}).  Given a total desorption rate $R_{\mathrm{des}} = \Sigma_i(R_{\mathrm{tdes},i} + R_{\mathrm{pdes},i}$), the rate of filling of a species $i$ from the Layer 2 mantle to the surface is given by
\begin{equation}
\label{eq:R_repl}
    R_{\mathrm{repl},i,2-S} = -R{_\mathrm{des}}\frac{N_{i,2}}{{N_{2}}},
\end{equation}
where $N_{i,2}$ is the column density of species $i$ in the Layer 2 mantle, and $N_2$ is the total column of the Layer 2 mantle.  Layer 2 is in turn replenished from Layer 1:
\begin{equation}
        R_{\mathrm{repl},i,1-2} = -R{_\mathrm{des}}\frac{N_{i,1}}{{N_{1}}}
\end{equation}

Once Layer 2 is completely lost, the surface is replenished following Equation \ref{eq:R_repl} but from Layer 1.  In this way, desorption is allowed to occur until all the ice is lost.  Note that we do not explicitly treat diffusion between the layers, however as a result of desorption, material from lower layers is effectively mixed upwards within the ice.  This is done in a statistical way and does not account for the different diffusion barriers of different molecules, which in reality will dictate their mixing efficiency within the ice.  The implications of this are discussed in Section \ref{subsec:model_lim}.

\subsubsection{Photodissociation}
\label{sssec:photodissociation}
UV photodissociation cross-sections for each molecule, $\sigma_{\mathrm{pdiss},i}$, are taken from the Leiden Observatory database\footnote{https://home.strw.leidenuniv.nl/~ewine/photo/} \citep{Heays2017}.  In the ice phase, photodissociation products are trapped in place and can rapidly react to reform the parent molecule, lowering the effective photodissociation cross-section compared to the gas phase \citep[e.g.][]{Oberg2016}.  We therefore scale down the gas-phase cross-sections by a factor of 10 for use in our ice model.  The photodissociation rate constants are then calculated by
\begin{equation}
\label{eq:pdiss}
    k_{\mathrm{pdiss},i} = \sum_n \bar{\sigma}_{\mathrm{pdiss},i,n} \bar{F}_{UV,n}/2,
\end{equation}
where $n$ denotes the discrete wavelength bins sampled in our model, $\bar{\sigma}_{\mathrm{pdiss},i,n}$ is the average photodissociation rate constant within wavelength bin $n$ for species $i$, and $\bar{F}_{UV,n}$/2 is the geometry-corrected UV flux in wavelength bin $n$.  As noted in Section \ref{subsec:model_structure}, the wavelength bins are chosen such that the wavelength-summed photodissociation rate constant for all molecules (e.g. Equation \ref{eq:pdiss}) is within 5\% of the wavelength-integrated quantity $\int_{\nu=91\mathrm{nm}}^{200\mathrm{nm}}\sigma_{\mathrm{pdiss},\nu}  F_{UV, \nu}$, given the input stellar flux profile.

Because UV penetration is attenuated with ice depth, the photodissociation rate constant is adjusted in each monolayer:
\begin{equation}
    k_{\mathrm{pdiss},i,j} = k_{\mathrm{pdiss},i}(1-0.007)^{j-1},
\end{equation}
where $j$ is the depth of a given monolayer ranging from 1 to the total ice thickness $N_\mathrm{{monolayers}}$.  The factor 0.007 represents the probability of photon absorption within a single ice monolayer, taken from \citet{Andersson2008}.  Lastly, the mean photodissociation rate within each layer (e.g.~surface layer, layer 2 mantle, and layer 1 mantle) is found from: 
\begin{equation}
\label{eq:uv_atten}
    R_{\mathrm{pdiss},i} = \sum_{j_\mathrm{top}}^{j_{\mathrm{bottom}}} k_{\mathrm{pdiss},i,j} N_i/n_{\mathrm{lay}},
\end{equation}
where $j_\mathrm{top}$ and $j_\mathrm{bottom}$ denote the layer boundaries in terms of monolayers from the surface, resulting in a total layer thickness $n_\mathrm{lay} = j_\mathrm{bottom} - j_\mathrm{top}$.  Thus, $N_i/n_\mathrm{lay}$ represents the total column density of species $i$ distributed across $n_{\mathrm{lay}}$ monolayers within a given ice layer.

We track dissociation products as distinct species from the parent molecules within our ice model.  In subsequent analysis, we will refer to the parent molecules and dissociation products as `primary' and `secondary' ices, respectively.  For dissociations within the surface layer, the secondary species are treated as a single product with the binding energy of the heavier radical product (Table \ref{tab:E_bind}).  For dissociations within the mantle layers, we treat the secondary species as a single product with the same binding energy of the original parent molecule.  This treatment assumes that radicals within the mantle are likely to recombine with neighboring radicals to form molecules with a binding energy closer to the parent molecule than to the radical products.  In this way, we capture the salient effects of radical recombination within the ice without needing to implement a detailed treatment of ice diffusion or formation chemistry.  Indeed, since thermal desorption plays only a minor role in most disk regions (Section \ref{subsec:static_destr}), and photodesorption rates are uniform across all molecules, the identity of the recombination product should not meaningfully impact the ice survival results.  While secondary ices are considered in the accounting of ice thickness used to attenuate the incident UV flux (Equation \ref{eq:uv_atten}), we do not explicitly track any further dissociation of secondary ices.

\begin{figure*}
    \includegraphics[width=\linewidth]{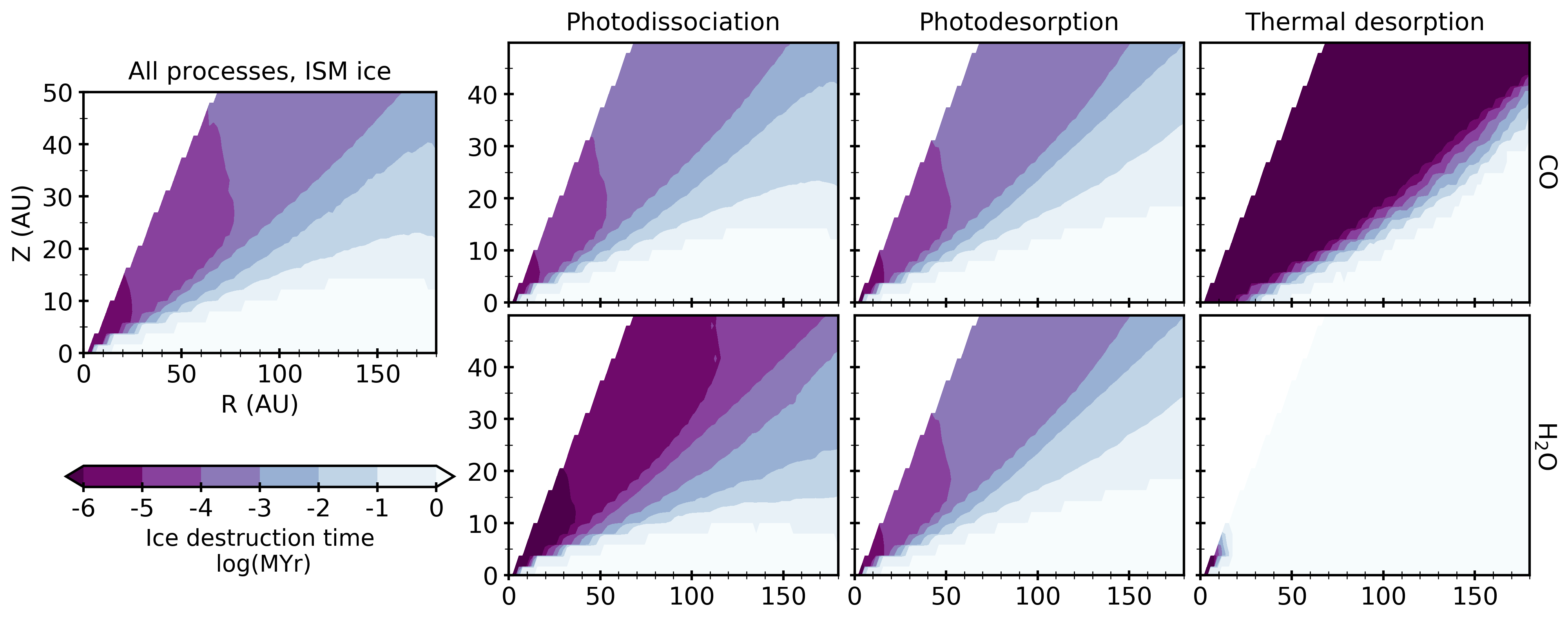}
    \caption{Left: times required for total loss of the primary ice mantle at different positions within a static disk, considering the ISM ice structure described in Section \ref{subsec:model_chem} and including photodissociation, photodesorption, and thermal desorption.  Right: ice destruction times for individual destruction processes, considering a 100 ML ice of pure CO (top) or pure H$_2$O (bottom).}
    \label{fig:static_mech}
\end{figure*}

\subsection{Chemical model caveats}
\label{subsec:model_lim}
Our ice chemistry model is designed to capture the important microphysics regulating ice survival, while still maintaining computational feasibility needed to run thousands of trajectories.  We do not consider chemical processing due to exposure to X-rays and cosmic rays, which should penetrate closer to the midplane than UV.  This is motivated by detailed modeling that shows photochemistry timescales are $>$10$^6$ years in the disk midplane even when considering X-ray and cosmic ray induced chemistry along with UV \citep{Semenov2011}.  Therefore, on the timescales considered for our analysis, we do not expect the inclusion of X-rays or cosmic rays to significantly increase ice destruction in the midplane.

We also do not include ice diffusion in our model, though some upward diffusion takes place via the replenishment of the desorption surface by mantle molecules.  Because we do not explicitly treat reactions within our ices, it is not important to include diffusion to properly track the chemistry.  The main impact of omitting diffusion in our models is that we likely overestimate the extent to which CO (and perhaps also CH$_4$) is trapped in the ice mantle within its snow line.  When particles drift interior to the CO snowline, they typically lose about 20\% of the original CO to thermal desorption, at which point the desorption surface becomes saturated with less volatile molecules.  Laboratory experiments of astrophysical ice analogs show CO trapping efficiencies up to $\sim$10\% in H$_2$O ice and 65\% in CO$_2$ ice \citep{Simon2019}.  We therefore expect that the true extent of CO loss should be a few times higher than our models predict.  Still, this overestimation of CO trapping does not meaningfully impact any of our conclusions.

That we do not include formation chemistry in our models should also not impact the inferred ice survival outcomes.  Given that neutral-neutral reactions typically have energy barriers, the majority of ice-phase chemistry should be driven by the production of radicals.  In high-density regions, we also expect limited hydrogenation chemistry as most H should be in the form of H$_2$.  Thus, chemistry should not significantly contribute to the destruction of ice species that are not already destroyed via photodissociation.  Still, understanding the formation of organics in these dynamic ice particles is an interesting avenue for predicting the inventories of prebiotic organics synthesized in situ in disks.  A full treatment of ice-phase chemistry will be the subject of future work.

Lastly, we neglect gas-phase chemistry and, in particular, the freeze-out of gas-phase molecules onto grains.  CO should be the most abundant molecule after H$_2$ in disks.  Interior to the CO snow surface, CO should not efficiently adsorb onto the grains.  Exterior to the CO snow surface, the gas-phase CO abundance will be extremely low, and further freeze-out of CO onto grains should not be important.  Moreover, photodissociation is the dominant destruction mechanism in most disk regions of our models.  While photodissociation rates are attenuated with ice depth, the dependence on ice thickness is rather small, scaling as a factor of $(1-0.007)^{j-1}$ where $j$ is the depth of a given monolayer (Section \ref{sssec:photodissociation}).  Thus, even accumulating 100 ML of new ice would only attenuate the photodissociation rates of the original ice by a factor of $\sim$2.  Accretion of new ice will inhibit photodesorption and thermal desorption, but these play only a small role in ice loss in most of the disk.

\subsection{Ice destruction in a static disk}
\label{subsec:static_destr}
As a first look at ice destruction in the disk environment, Figure \ref{fig:static_mech} shows ice loss timescales for a static disk.  The left panel shows the time required for the entire primary ice to be destroyed when considering all destruction processes simultaneously.  It is clear that ices in the disk midplane are well shielded from destructive processing.  The vertical gradients in ice loss timescales are quite steep, especially in the inner $\sim$50 au, and so ice destruction becomes orders of magnitude more efficient over a relatively small change in elevation.   

The right panels of Figure \ref{fig:static_mech} shows the ice loss timescales for each destruction process individually, considering either a pure H$_2$O ice or a pure CO ice.  These molecules are illustrative because they exhibit opposite destruction behavior: CO experiences efficient thermal desorption but inefficient photodissociation, while H$_2$O experiences inefficient thermal desorption and more efficient photodissociation.  These plots therefore demonstrate the range of possible outcomes for different ice destruction mechanisms.  Note that photodesorption efficiencies are uniform for all molecules in our models. 

Most molecules are not efficiently thermally desorbed in the disk regions considered here ($>$4 au; Section \ref{subsec:chem_dyn_model}).  CO is a notable exception, with a midplane snowline around 20 au.  CO is also somewhat unusual in that its photodissociation rate is low compared to the other molecules in our simulation.  The CO photodissociation efficiency is therefore comparable to its photodesorption efficiency.  For other molecules, photodissociation is typically much more efficient than photodesorption, as illustrated by H$_2$O.  Therefore, we expect photodissociation to be the most important mechanism of ice destruction in our model.  Photoprocesses are generally not efficient below an elevation $z/r \sim$0.1-0.2 in our model.  Dynamic cycling of particles into more elevated, UV-exposed disk layers is needed to drive any meaningful ice loss (besides thermal desorption of hypervolatiles) at the disk radii considered in this work. 

\subsection{Chemodynamic modeling}
\label{subsec:chem_dyn_model}

To calculate ice destruction for individual particle trajectories, at each time step we find the local temperature and UV flux at the particle's position using a 2D linear interpolation of the disk temperature and UV structures solved for in Section \ref{subsec:model_structure}.  Ice destruction rates are then calculated following the formalisms in Sections \ref{sssec:tdes}--\ref{sssec:photodissociation}.  The calculation is terminated when either (i) all primary and secondary ices are lost from the grain, (ii) the particle reaches an inner radius of 4 au, or (iii) the maximum run time for the trajectory is met.  We choose a cutoff radius of 4 au because this allows us to probe the minimum extent of the comet formation zone \citep{Bockelee-Morvan2004}, while maintaining manageable computational costs due to the decreasing time steps required in the inner disk.  

\begin{figure*}
    \includegraphics[width=\linewidth]{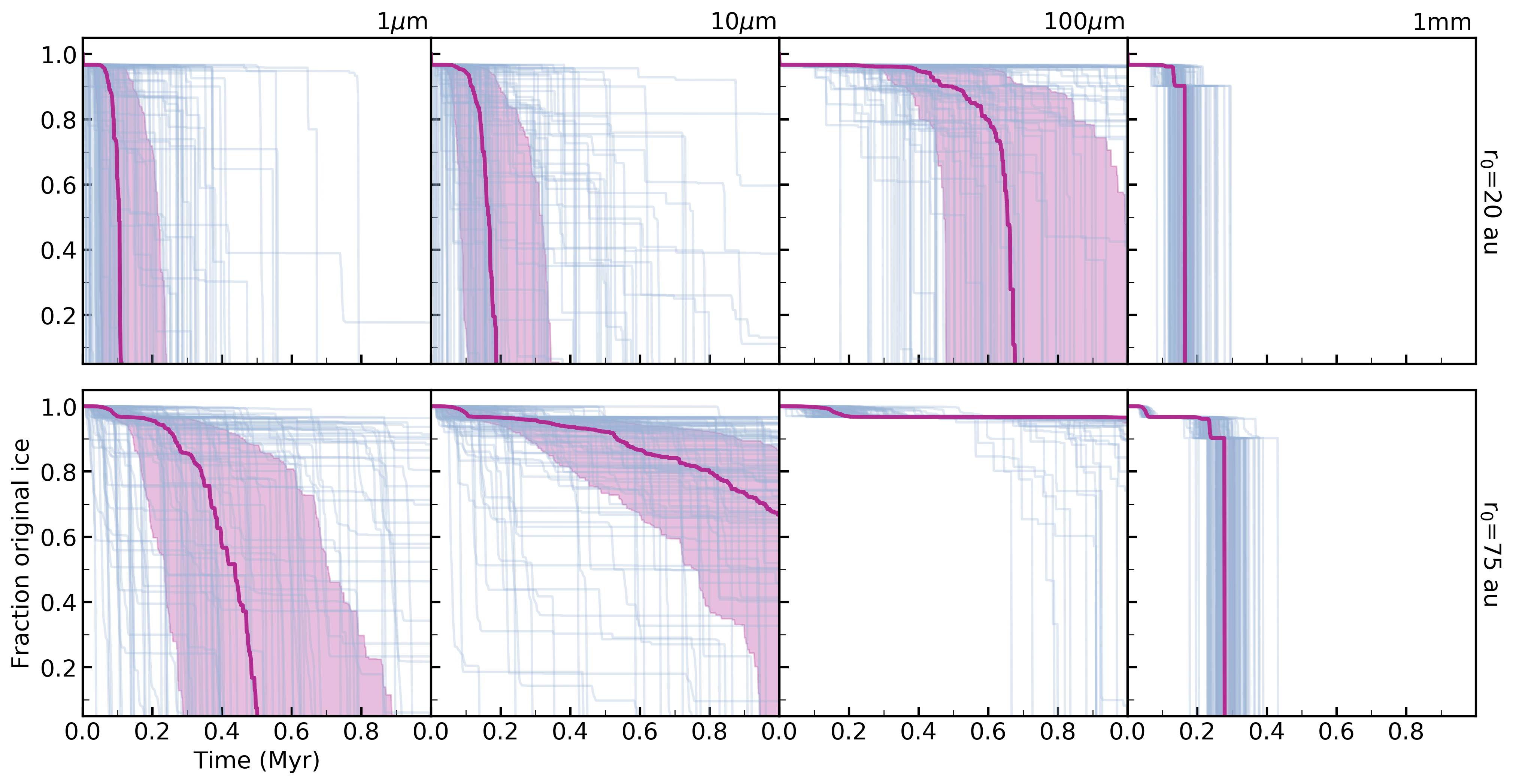}
    \caption{Ice survival fraction (primary + secondary ice) for particle trajectories as a function of time for 1 Myr runs.  Individual trajectories are shown as blue lines.  Pink lines and shaded regions represent the median survival fraction and the 25--75th percentile range, respectively, of all trajectories.}
    \label{fig:fid_loss_time}
\end{figure*}

We run batches of models for both long (1 Myr) and short (100 kyr) timescales.  For the long group, we run 250 trajectories for each grain size, for particles initialized at both 20 and 75 au.  For the short group, we run 100 trajectories for each grain size, for particles initialized at 20, 50, 75, 100, and 150 au.  The long trajectories are likely not physical, as grain growth is expected to happen on faster timescales (Appendix \ref{sec:app_growth_times}).  However, these trajectories allow us to evaluate the characteristic behavior of ice destruction in different disk regions for different grain sizes.  The short trajectories are more physical, and we explore a wider range of parameter space in order to infer plausible ice destruction outcomes.  In the following Section, we present the ice destruction results for these models.

\section{Results}
\label{sec:results}

\subsection{Ice destruction timescales: 1 Myr models}
\label{subsec:ice_destr_timescales}

We first consider the results of our long (1 Myr) model suite in order to evaluate the characteristic ice destruction timescales for different particle sizes and disk radii.  Figure \ref{fig:fid_loss_time} shows the fraction of the original ice mantle (primary + secondary ice species) remaining as a function of time for individual particle trajectories, along with the median and 25-75th percentiles for ice survival across all particles with a given size and initial radius.  

Small particles (1-10$\mu$m) are subject to significant ice destruction, particularly at smaller radii in the disk.  As can be seen in the trajectory clouds of Figure \ref{fig:traj_cloud}, particles in this regime are readily lofted to elevated disk layers and subject to UV exposure.  The ice loss timescale for small grains therefore primarily represents the timescale for a particle to be cycled into unshielded disk layers, which is most efficient for smaller particles at smaller radii.  For 1mm sized grains, the particles remain settled in the shielded midplane for a few hundred kyr and experience minimal ice destruction during this time.  However, these grains are drifting into the star, and experience some ice loss due to thermal desorption of CO around 20 au, and HCN and H$_2$CO around 7 au.  We assume that all ice is lost when they reach 4 au, the inner boundary of our simulation (Section \ref{subsec:chem_dyn_model}).  100$\mu$m sized grains represent an interesting intermediate case: particles initialized at 20 au are susceptible to both drift and vertical cycling, resulting in considerable ice destruction though with typically long destruction timescales ($>$0.5 Myr).  At larger radii, we see almost no ice destruction for 100$\mu$m grains since the particles are large enough to avoid cycling into the unshielded layers, but small enough to avoid drift on 1 Myr timescales.

\begin{deluxetable*}{lcccc}
	\tabletypesize{\footnotesize}
	\tablecaption{Ice loss timescales (Myr) \label{tab:loss_times}}
	\tablecolumns{5} 
	\tablewidth{\textwidth} 
	\tablehead{
		\colhead{$r(0)$ (au)}       & 
		\colhead{1 $\mu$m} &
		\colhead{10 $\mu$m} &
		\colhead{100 $\mu$m} &
	    \colhead{1 mm} }
\startdata
\multicolumn{5}{c}{Primary + secondary ices} \\
\hline
20 au & 0.10 [0.05, 0.22] & 0.17 [0.08, 0.32] & 0.66 [0.47, $>$1] & 0.16 [0.14, 0.19]\\
75 au & 0.44 [0.24, 0.70] & $>$1 [0.76, $>$1] & $>$1 [$>$1, $>$1] & 0.28 [0.25, 0.31]\\
\hline
\multicolumn{5}{c}{Primary ices} \\
\hline
20 au & 0.07 [0.03, 0.13] & 0.10 [0.05, 0.17] & 0.52 [0.36, 0.78] & 0.16 [0.14, 0.19]\\
75 au & 0.21 [0.10, 0.33] & 0.31 [0.15, 0.55] & $>$1 [$>$1, $>$1] & 0.28 [0.25, 0.31]\\
\enddata
\tablenotetext{}{Listed values represent the time when 50\% of the ice was lost for the 50th [25th, 75th] percentile particle in each group.  Lower limits are listed when 50\% of ice was not lost on a 1 Myr timescale.}
\end{deluxetable*}

Table \ref{tab:loss_times} lists the time at which the 25, 50, and 75th percentile particles within each simulation group have lost 50\% of the ice mantle.  This metric captures the relative efficacy of ice destruction between different grain sizes and disk radii, as well as the spread in destruction timescales within a given simulation group.  We consider the timescales for both the loss of primary + secondary ice, as well as the loss of the primary ice species alone.  

The median ice loss timescales are generally a few hundred kyr or longer.  Here, it becomes important to consider other processes that are taking place in the disk on similar timescales.  As demonstrated in Appendix \ref{sec:app_growth_times}, grain growth will take place on timescales of a few to tens of kyr in the disk midplane.  Thus, while the 1 Myr trajectories are useful from the standpoint of deriving characteristic ice destruction timescales, they are likely not physical.  For subsequent analysis, we focus on timescales of 100 kyr in order to conservatively encompass the time that a particle is likely to remain a given size.

\subsection{Survival outcomes: 100 kyr models}

We now consider in detail the survival of ice species for a more realistic timescale of 100 kyr.  We consider initial radii of 20, 50, 75, 100 and 150 au in order to sample ice survival outcomes across the disk.  100 trajectories are run for each combination of grain size and initial radius, since we found that this is enough to recover the statistical trends found from 250 trajectories.  

\begin{figure*}
    \includegraphics[width=\linewidth]{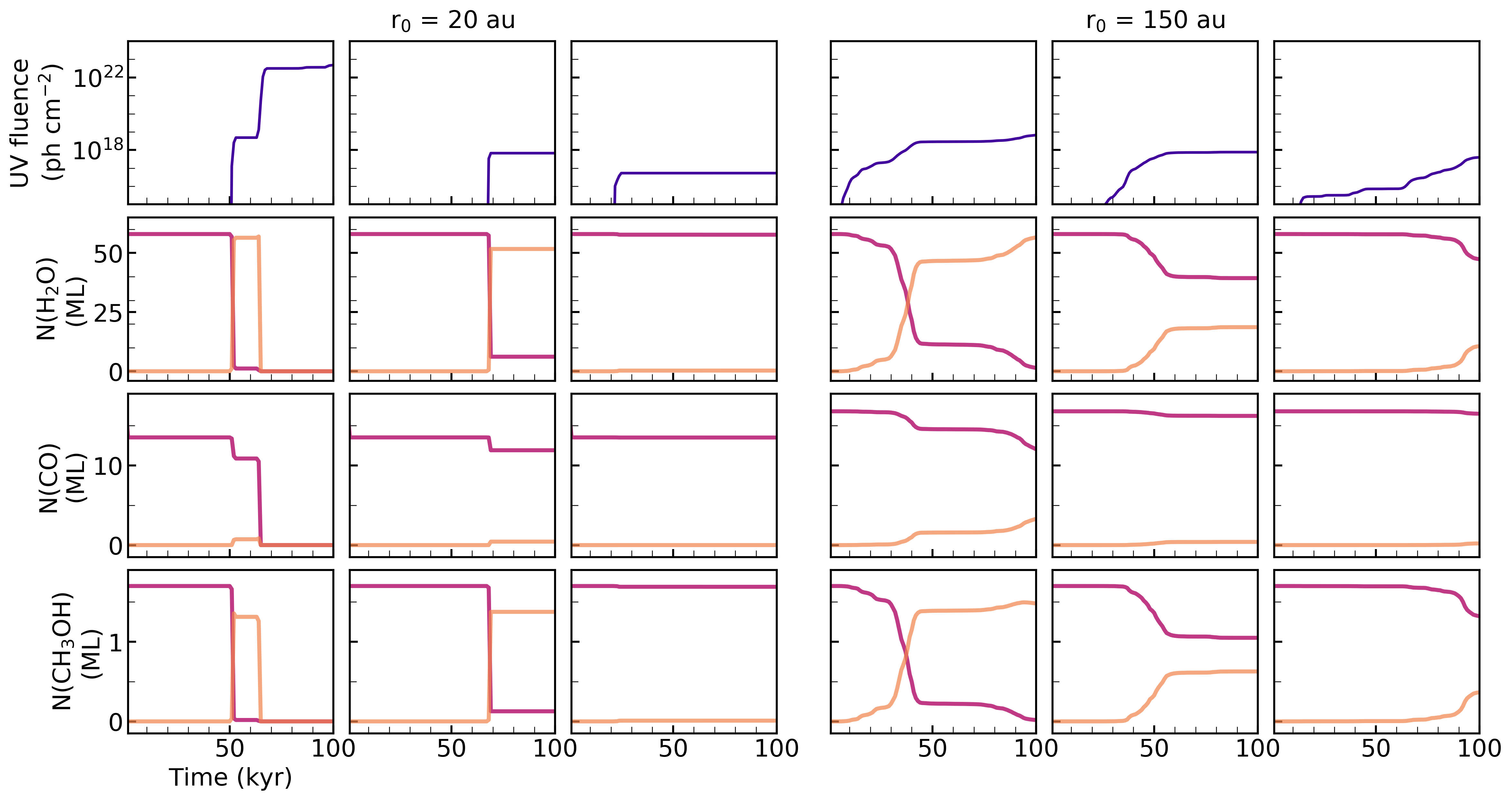}
    \caption{Three example trajectories for 1$\mu$m grains starting at 20 au (left) and 150 au (right).  The UV fluence encountered by the particle is shown in the top row.  The ice mantle inventories of H$_2$O, CO, and CH$_3$OH are shown in the bottom rows.  Pink lines represent the primary ice species, and orange lines represent the secondary species (i.e.~dissociation products).}
    \label{fig:ex_comp_H2O_CO}
\end{figure*}

\subsubsection{Ice destruction trends}
\label{sssec:destr_examples}
To illustrate the ice destruction behavior for individual particles, Figure \ref{fig:ex_comp_H2O_CO} shows three example particle trajectories for 1$\mu$m sized grains starting at both 20 au and 150 au.  There are three important trends to highlight: (i) the behavior of molecules with different photodissociation rates, (ii) the behavior of molecules in ice Layer 1 (H$_2$O-rich) vs. Layer 2 (CO-rich), and (iii) the behavior of grains at small vs. large disk radii.  

(i) Of the molecules considered in our model, most have similar photodissociation rates.  The exceptions are CO$_2$ and CO, which typically have photodissociation rates 1 and 2 orders of magnitude lower, respectively.  This is largely due to the fact that CO$_2$ and CO are not efficiently photodissociated by Ly-$\alpha$ photons, which carry the majority of the photon flux in the disk.  The difference in ice behavior between the fast photodissociation group (H$_2$O, CH$_4$, NH$_3$, CH$_3$OH, H$_2$CO, and HCN) and the slow photodissociation group (CO and CO$_2$) is illustrated by H$_2$O and CO in Figure \ref{fig:ex_comp_H2O_CO}.  H$_2$O is rapidly converted to secondary ice species upon exposure to high UV fields, while CO is more resilient to photodissociation.  Thus, we expect CO and CO$_2$ to remain intact as primary ices more readily than the other molecules in our simulation.

(ii)  The difference between ice species in Layer 1 vs. Layer 2 is illustrated by H$_2$O and CH$_3$OH in Figure \ref{fig:ex_comp_H2O_CO}.  While H$_2$O is readily photodissociated upon exposure to a strong UV field, the total inventory of primary + secondary H$_2$O is generally conserved until a strong enough UV field is encountered that the ice is completely photo-desorbed.  In contrast, CH$_3$OH is simultaneously photodissociated and photodesorbed since it occupies the desorptive surface layer.  CO presents an intermediate case, as it is present in both Layer 1 and Layer 2.  Thus, we expect molecules present primarily in the CO-rich ice layer to be preferentially lost from the ice, either as the primary species or its dissociation product.  Similarly, molecules present in the H$_2$O-rich layer should be preferentially preserved, though in some cases entirely in the form of the dissociation product.

(iii) The vertical gradients in temperature and UV flux are steeper in the inner disk than the outer disk.  As a result, particles cycled vertically will experience pulses of ice processing, illustrated by the $r_0$=20 au grains in Figure \ref{fig:ex_comp_H2O_CO}.  These pulses are generally quite destructive when they occur, resulting in steep drops in primary ice column densities or even total loss of the ice mantle.  On the other hand, the reduced shielding in the outer disk translates to a more persistent exposure to UV radiation, but at lower levels than those experienced in the inner disk.  As a result, ice destruction takes place gradually over the entire particle trajectory.  Moreover, the lower UV fluxes in the outer disk result in low photodesorption rates, meaning that secondary ice species generally accumulate within the mantle.  

\begin{figure*}
    \includegraphics[width=\linewidth]{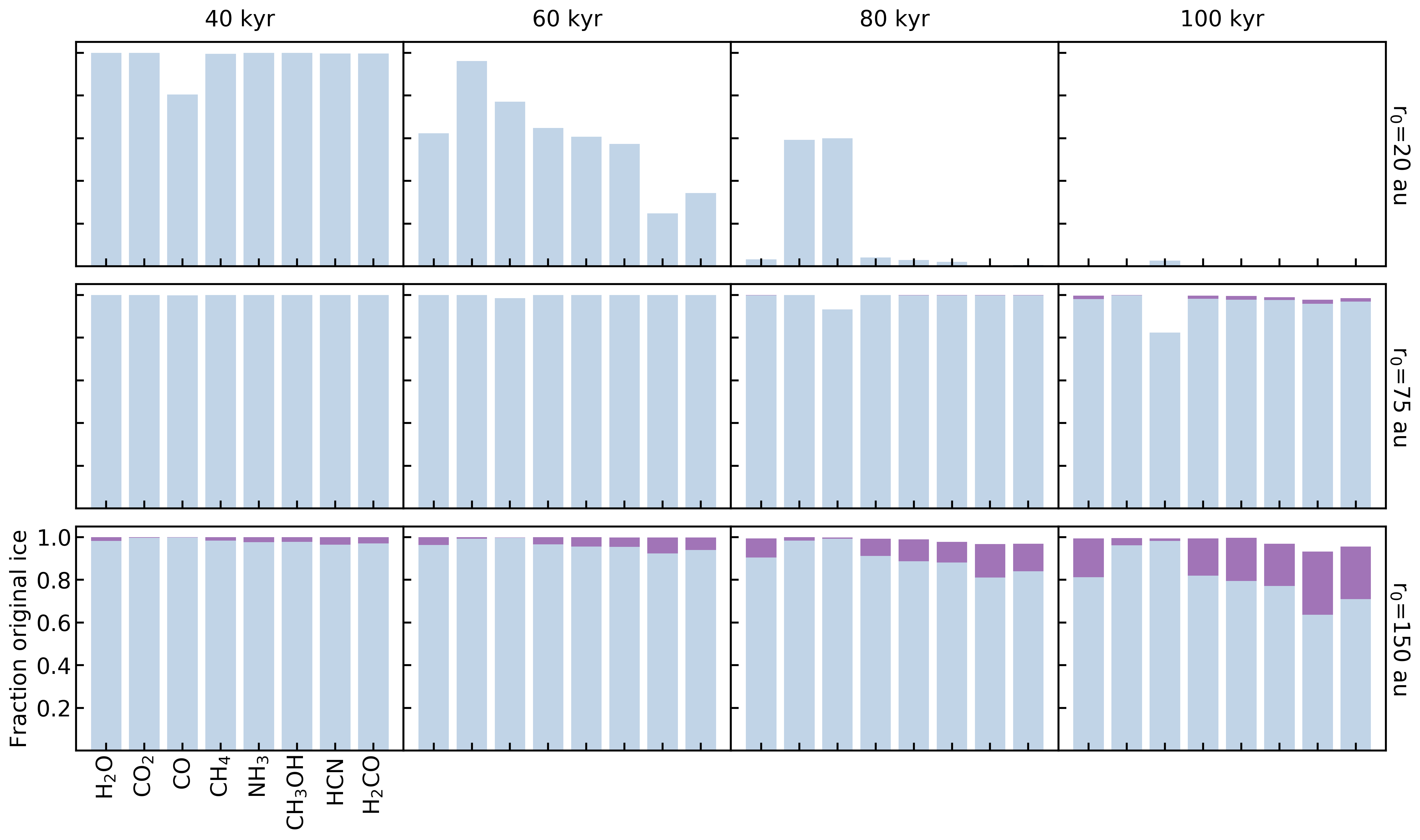}
    \caption{Median ice survival fractions for 1$\mu$m grains at different time points in the simulation, shown for initial radii of 20, 75, and 150 au.  Blue and purple bars represent primary and secondary ices, respectively.}
    \label{fig:species_v_time}
\end{figure*}

With these behaviors in mind, we now consider the ice survival outcomes in a more statistical sense, combining the results from all trajectories within a given simulation group.  

\subsubsection{1$\mu$m ice destruction statistics}
\label{sssec:destr_1mic}

We first focus on 1$\mu$m sized grains since they exhibit the most ice destruction of any grain size on a 100 kyr timescale, and therefore allow us to identify trends in ice destruction behavior.  Figure \ref{fig:species_v_time} shows the median (across 100 trajectories) fraction of each ice species surviving at four different time points in the simulation, for 1$\mu$m grains initialized at 20, 75, and 150 au.  Primary and secondary species are represented with blue and purple bars, respectively. 

Particles initialized at 150 au show photodissociation of primary ices by a few tens of kyr.  Several of the trends noted in Section \ref{sssec:destr_examples} are seen to play out in the statistics we observe here.  First, we note that the easily photodissociated molecules (H$_2$O, CH$_4$, NH$_3$, CH$_3$OH, HCN, and H$_2$CO) show much greater loss than CO and CO$_2$. Moreover, molecules that occupy only the CO-rich layer show more loss than molecules that occupy the H$_2$O-rich layer (e.g.~CH$_3$OH, HCN, and H$_2$CO vs.~H$_2$O, NH$_3$, and CH$_4$).  This reflects that molecules in the lower ice layers cannot be desorbed, and also are photodissociated at a slightly slower rate due to shielding with increased ice depth.  For 150 au particles, secondary ices make up a meaningful contribution to the median ice composition due to the steady production of secondary ices in the outer disk discussed in Section \ref{sssec:destr_examples}.  Photodesorption rates are also low in the outer disk, meaning that secondary species accumulate on the grain surfaces.

For 20 au particles, the median ice composition shows minimal ice loss prior to 60 kyr.  The exception is CO, for which $\sim$20\% of the original ice is lost due to thermal desorption in the beginning of the simulation.  CO desorption ceases when the surface is filled with less volatile molecules, trapping most of the CO ice in the mantle.  As described in Section \ref{subsec:model_lim}, the true extent of CO loss is likely higher than our models predict, but our results are not meaningfully impacted by this.  At 60 kyr, the median ice composition has changed significantly as particles have had time to diffuse into more elevated disk layers.  The relative photodissociation efficiency of different ice species follows a similar pattern as noted for ice loss at 150 au: ice species with high photodissociation rates are destroyed most efficiently, especially those in the CO-rich layer.  CO and CO$_2$ are present in both the CO-rich and H$_2$O-rich layers but are less susceptible to photodissociation and so are not as efficiently destroyed.  In addition, thermal desorption also contributes to ice loss at 20 au.  This is apparent in the enhanced loss of HCN and H$_2$CO at 20 au relative to 150 au, which is due to a combination of desorption of the photodissociation products from the surface layer of the ice and, in warmer ($\gtrsim$30 K) regions, direct desorption of the parent molecules.  Only molecules occupying the CO-rich ice layer are impacted by thermal desorption, while similarly volatile molecules or radicals in the H$_2$O-rich layer remain trapped in the mantle.  We note that the HCN and H$_2$CO binding energies adopted here are low compared to laboratory measurements (see Section \ref{sssec:tdes}), and so our models may overpredict the true extent of parent molecule thermal desorption.

At 20 au, there is no appreciable contribution of secondary species to the total ice when considering the median column densities across all trajectories.  This is likely due to the transient and stochastic nature of secondary ice production at small disk radii.  Here, particles entering the unshielded disk layers undergo rapid production of secondary ices but often fully lose the ice mantle shortly thereafter.  Additionally, some particles never enter the unshielded region, and do not produce secondary species in the first place.  As a result, at any given time step, the median column densities of the secondary ice species are negligible.  We therefore emphasize that the lack of secondary ice for the 20 au simulations is a statistical representation and should not be interpreted as reflecting the ice composition of any single particle.  Thus, individual particles never exhibit only CO and CO$_2$ without any other primary or secondary ices, as might be inferred at 80 kyr.  Instead, CO and CO$_2$ are the most likely molecules to survive intact, and should regularly be found intact on ice mantles up to $\sim$80 kyr.  Similarly, while secondary ices are produced in abundance in individual trajectories (Figure \ref{fig:ex_comp_H2O_CO}), it is statistically unlikely that they will be present on a given particle at a given time step.

The 75 au particles show little ice destruction until 100 kyr, except for thermal desorption of CO as particles reach moderately elevated regions with temperatures above $\sim$20 K.  Interestingly, this intermediate disk radius exhibits much less ice destruction than the 20 au or 150 au cases.  The vertical gradient in the UV field is shallower at 75 au compared to the inner disk, and thus in contrast to the 20 au models, particles are not easily lofted to the disk elevations required for UV photoprocessing.  The higher densities at 75 au compared to 150 au result in more effective UV shielding, diminishing ice destruction relative to the 150 au models.  Thus, intermediate disk regions experience very little ice processing even for 1$\mu$m sized particles, as it takes $>$100 kyr for a typical grain to diffuse into a disk region that is UV-exposed (Figure \ref{fig:fid_loss_time}).

\subsubsection{Ice destruction for different grain sizes}
\label{sssec:destr_allsizes}

\begin{figure*}
    \includegraphics[width=0.9\linewidth]{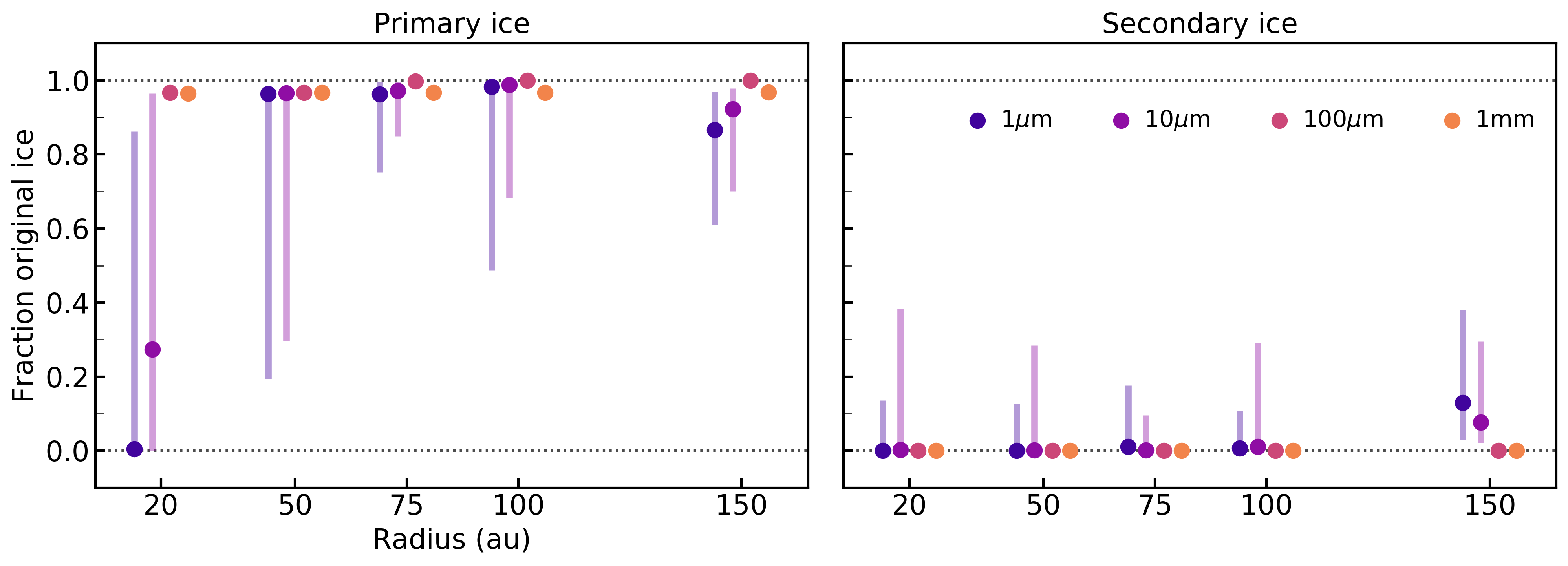}
    \caption{Summary of ice survival statistics at 100 kyr for primary and secondary ice species with respect to the initial ice thickness.  For each combination of grain sizes and initial radii, scatter points represent the median survival outcome and shaded bars show the 25-75th percentile range.}
    \label{fig:survival_summary}
\end{figure*}

We now synthesize the ice survival outcomes for all modeled grain sizes (1$\mu$m--1mm) and disk radii (20--150 au).  Figure \ref{fig:survival_summary} summarizes the ice loss statistics, where all primary and secondary ice species have been grouped together.  The relative survival of individual molecules generally reflects the same trends seen for the 1$\mu$m grains discussed in Section \ref{sssec:destr_1mic}.  

Of the particle sizes considered in our model, the 1$\mu$m particles exhibit the most significant ice destruction due to the relative ease of lofting these grains into the exposed upper disk layers.  10$\mu$m grains are also susceptible to ice loss, and show similar trends to the 1$\mu$m grains though to a lesser extent.  Notably, the loss of primary ice species is most efficient at small radii (20 au) and large radii (150 au).  While 20 au and 150 au are the only disk radii where the median particle shows more than a few percent ice loss, all radii show tens of percent of ice loss at the 25th percentile level.  Thus, a minority but appreciable population of 1 and 10$\mu$m sized particles will experience moderate ice destruction at all disk radii on a 100 kyr timescale.  For both small grain sizes, we also see that secondary ices typically accumulate to a greater extent at 150 au compared to the smaller disk radii.  Again though, the 25th percentile reflects that for all disk radii, tens of percent of the ice will be retained as secondary ices for a subset of the population.

The larger grain sizes, 100$\mu$m and 1mm, show almost no ice loss by 100 kyr.  The few percent of ice that is lost reflects the thermal desorption of volatile species from particles that drift to lukewarm ($\sim$20--30 K) disk radii.  This affects all 1mm sized particles along with 100$\mu$m particles initialized within 50 au.  Because this is a thermal process, secondary ice species are not produced on these larger grains.  We see no increase in ice destruction at the 25th percentile level, indicating that ice destruction is generally not an important outcome for these larger grains.  On longer timescales, drift (1mm and 100 $\mu$m) and vertical diffusion (100 $\mu$m) will begin to result in further ice loss, as discussed in Section \ref{subsec:ice_destr_timescales}.  However, these processes do not occur quickly enough to meaningfully impact ice survival on realistic timescales.

\section{Discussion}
\label{sec:disc}

\subsection{Ice inheritance and comet formation}
\label{ssec:comet_formation}
In our models of ice destruction for particles undergoing diffusion, settling, and drift within a disk, ice destruction is relatively inefficient on 100 kyr timescales, except for small grains ($<$10$\mu$m) in the inner or outer disk (e.g.~20 and 150 au in our models; Figure \ref{fig:survival_summary}).  With this in mind, we now assemble a plausible scenario for how interstellar ice is incorporated into icy bodies in the outer solar system.  It is important to recognize that this interpretation assumes our adopted disk physical structure is a good analog for the Solar Nebula.  The impacts of various disk physical properties on ice survival outcomes are discussed further in Section \ref{ssec:disc_caveats}.

Protostellar infall models indicate that pristine ice is preferentially incorporated into the outer disk regions \citep{Visser2011, Drozdovskaya2014}.  Grains should enter the disk with small sizes ($\mu$m scale).  At 150 au, 1--10$\mu$m grains will typically experience some degree of ice destruction during the time spent growing to $\sim$100$\mu$m in size.  While particles can lose up to tens of percent of their pristine ices to photodissociation in this regime, the dissociation products are not efficiently desorbed beyond a few tens of au, and should generally remain in the ice as secondary species (Figure \ref{fig:survival_summary}).  It is also important to note that there is a wide range in the degree of ice processing experienced by small particles, and therefore growing particles may incorporate a mixture of unprocessed and moderately processed ices.  

Once grains reach sizes of $\sim$100$\mu$m, they are not circulated to the UV-rich, upper disk regions and so further ice destruction is inefficient.  Following further growth to millimeter sizes, grains are transported inwards towards the star (Figure \ref{fig:traj_cloud}).  These drifting particles should contain contributions of unprocessed and processed (i.e.~photodissociated) ice species inherited from smaller grains, but should not experience further ice destruction.  The exception is that some hypervolatiles will be lost once drifting grains pass interior to their snowlines.  While our model likely under-estimates the extent of hypervolatile loss, we still expect tens of percent of hypervolatiles to stay trapped in the ice exterior to the H$_2$O and CO$_2$ snowlines (see Section \ref{subsec:model_lim}).  Icy bodies such as comets that accrete these drifting pebbles should therefore incorporate significant reservoirs of pristine ice, with the majority in the form of the primary ice species and up to a few tens of percent in the form of radical recombination products.  In either case, interstellar isotopic signatures should be preserved.  

Thus, the assembly of cometary bodies via accretion of drifting pebbles formed in the outer disk can explain the incorporation of ices with interstellar compositions and isotopic ratios.  Pebble accretion fed by drift from the outer disk has been shown to explain the production of gas-giant and terrestrial planets, and there is evidence that such a process is at play in the assembly of Solar System bodies \citep[e.g.][]{Lambrechts2014, Lambrechts2019, Johansen2021}.  Still, more detailed modeling is needed to address whether it can also explain the properties and demographics of comets.

An alternative model for comet formation is if they are assembled from local material at their formation location within the disk, through hierarchical growth and/or the streaming instability \citep{Weidenschilling1997, Davidsson2016, Lorek2018}.  In the Solar System, the comet formation zone is estimated to be $\sim$5--30 au \citep{Bockelee-Morvan2004}.  Assuming that pristine ices survived the disk formation process at these radii, 1 and 10 $\mu$m sized grains should subsequently experience significant destruction of their icy reservoirs on timescales comparable to grain growth timescales (Figure \ref{fig:survival_summary}).  Indeed, at 20 au ices are not just photodissociated but are fully lost from the grain.  Thus, beginning with a population of exclusively small particles, we would expect minimal incorporation of pristine material into larger grains as they are forming.  Local comet assembly therefore cannot easily explain the incorporation of pristine icy material.  Under the assumption that this framework is applicable to the Solar Nebula, our models favor a scenario in which comets formed with a contribution of drifting icy pebbles from the outer disk.

It is interesting to consider how ice inheritance prospects depend on the disk evolutionary stage.  We can explore trends in ice survival as a function of disk evolution by varying the adopted large grain fraction $X_\mathrm{lg}$, which reflects the extent to which grain growth has proceeded in the disk.  This parameter strongly moderates the disk UV penetration as most UV opacity is due to small dust grains.  As seen in Appendix \ref{sec:time_dependence}, for a highly unevolved disk ice destruction is less efficient at 20 au compared to our fiducial model, but small grains still exhibit considerable destruction.  For instance, the median 1$\mu$m particle loses $\sim$80\% of its water by 100 kyr.  We therefore expect that even in young disks, local assembly of comets should not efficiently preserve interstellar signatures.  Beyond 20 au, there is virtually no ice destruction for any grain size, reflecting that ices in the outer disk are very well shielded and implying that preservation in these regions should be near-complete early in the disk lifetime.

In very evolved disks, ice destruction is much more efficient compared to our fiducial model: small ($<$10$\mu$m) grains lose most of their ice mantles at all disk radii, and 100$\mu$m grains lose at least half of their ices beyond 100 au.  1mm grains remain largely unaffected by ice loss, implying that ices that have already been incorporated into $>$1mm sized grains should be preserved even in a very evolved disk.  Thus, inheritance signatures within the Solar System likely correspond to material that was incorporated into larger grains fairly early in the disk lifetime, and remained confined to the midplane as the disk became increasingly UV-transparent.

\subsection{Extension to other disk structures} \label{ssec:disc_caveats}
We expect that various features of the disk physical structure will impact the ice survival outcomes.  A challenge to exploring this in a systematic way is that many disk properties are inter-related: for instance, the turbulence level in the disk will impact not only the efficiency of grain lofting to UV-exposed disk layers, but also the degree of dust settling \citep[e.g.][]{Dullemond2004} which impacts UV propagation.  In this regard, our disk model is not fully self-consistent, and a full exploration of parameter space is beyond the scope of this work.  Still, we highlight several qualitative changes to the survival outcomes that we expect given different disk physical structures, informed by test runs of disks with varying parameters.  Ice survival will increase in scenarios where UV propagation is reduced, including notably for an increased disk dust mass or for a star with a lower UV luminosity.  Similarly, ice loss should be reduced when grain lofting away from the midplane is inhibited, as in the case of a lower turbulent $\alpha$ or a decreased gas density.  Different radial and vertical density gradients will of course have a more complicated and location-specific impact on survival.  It is therefore likely that there is a range of inheritance outcomes for disks with different physical properties, and that the incorporation of pristine icy material proceeds to a greater or lesser extent in different planetary systems.  Disk and stellar properties are also likely to evolve temporally, which may contribute (along with grain growth, as discussed in Section \ref{ssec:comet_formation}) to different degrees of inheritance for bodies forming at different times.

\vspace{0.05in}
\subsection{Survival of isotopic carriers}

In comets, hydrogen and nitrogen show the largest variations in isotopic fractionation levels, and are therefore critical to inferring the origin of cometary ices \citep{Bockelee-Morvan2015}.  Some of the strongest evidence for interstellar inheritance in comets comes from high D/H ratios in H$_2$O, HCN, and CH$_3$OH; and high $^{15}$N/$^{14}$N ratios in HCN and NH$_3$, via NH$_2$ \citep{Altwegg2017, Drozdovskaya2021, Bockelee-Morvan2015}.  Additionally, $^{16}$O/$^{17}$O/$^{18}$O ratios in primitive meteorites are attributed to the preservation of fractionated H$_2$O inherited from the prestellar stage \citep{Yurimoto2004, Krot2020}.  Following the inheritance scenario outlined in the previous section, we find that it is certainly plausible that pristine reservoirs of H$_2$O, CH$_3$OH, HCN, and NH$_3$ can be transported to comet- and asteroid-forming disk radii via grain growth in the outer disk followed by pebble drift, at least exterior to their respective snow lines.

While there is a clear pathway to deliver interstellar isotopic signatures to icy planetesimals, there does not appear to be a straightforward relationship between the relative preservation of different molecules in ice grains and their fractionation levels within comets.  For instance, one could infer from our models that NH$_3$ should preserve interstellar $^{15}$N/$^{14}$N ratios better than HCN due to its higher resilience to destruction (Figure \ref{fig:species_v_time}).  In fact, the $^{15}$N/$^{14}$N ratios in these carriers are comparable to one another in comets, and in both cases much lower than ISM values \citep{Bockelee-Morvan2015}.  Given the relatively small differences in ice destruction between different molecules (factors of a few), the trends seen here are likely washed out in the general dilution of interstellar ice due to mixing with reprocessed material \citep[e.g.][]{Bockelee-Morvan2000, Brownlee2006, Rubin2019}.  Note also that gradients in the isotopic ratios within interstellar ice mantles, as predicted by \citet{Taquet2014} and \citet{Furuya2018}, may further complicate the inheritance of interstellar isotopic signatures, particularly if ice is preferentially lost from the surface.
 
Our models show that small (1-10$\mu$m) grains at disk radii beyond $\sim$20 au should experience efficient ice photodissociation but not desorption, resulting in the accumulation of secondary ice species.  This regime may be favorable for the production of complex organic molecules or even more refractory organic material upon recombination of radical fragments \citep[see also][]{Ciesla2012}.  This raises the possibility that interstellar-like isotopic signatures measured in meteoritic refractory organic material \citep[e.g.][]{Busemann2006} could originate from heavy processing of simpler ice species within the disk, rather than reflecting synthesis in the ISM.  Detailed chemical modeling is needed to more robustly explore the extent of in situ organic chemistry accompanying icy particle dynamics.

\subsection{Survival of prebiotically interesting molecules}

While our model does not consider large potentially prebiotic organics such as formamide and glycine, we can still outline rules of thumb for the inheritance of such species based on survival trends among simpler molecules.  We found that the relative destruction efficiency of a given ice molecule depends primarily on its UV photodissociation rate.  In general we see that molecules with high photodissociation cross-sections at Ly-$\alpha$ wavelengths typically have the highest overall photodissociation rates.  This should be broadly true when considering ice survival around T Tauri stars, for which Ly-$\alpha$ photons account for the majority of the total UV flux \citep{Schindhelm2012}.  

Of the organics in the Leiden Observatory photorates database \citep{Heays2017}, most complex organics (e.g.~CH$_3$CN, NH$_2$CHO, C$_2$H$_5$OH, CH$_3$CHO) have Ly-$\alpha$ photodissociation rates less than that of HCN, the ice species that was most efficiently photodissociated in our model.  We therefore expect that these molecules should survive at comparable levels to those shown in Figure \ref{fig:species_v_time} for molecules besides CO and CO$_2$.  For instance, 1$\mu$m grains at 150 au should preserve $\sim$60--80\% of their original organic inventory intact by 100 kyr.  However, we note that the larger organic C$_3$H$_7$OH exhibits a higher Ly-$\alpha$ photodissociation rate than HCN, and therefore has worse survival prospects.  Measurements of photodissociation cross-sections for additional large organics are required to better predict their survival outcome within ices.  

Ice survival will also depend on whether a molecule occupies the CO-rich layer only, or is also present in the H$_2$O-rich layer.  This is difficult to predict a priori for most large organics.  Still, in our model, molecules with comparable photodissociation rates that occupy different layers (e.g.~H$_2$CO vs.~NH$_3$) differ by just a few to a few tens of percent in their relative photodissociation levels.  We therefore expect that this effect should not dramatically alter survival outcomes, but will be secondary to the UV photodissociation efficiency.

\section{Conclusions}
\label{sec:concl}

We have developed a modeling framework to evaluate the prospects for interstellar icy material to survive passage through the disk and incorporation into icy bodies such as comets.  Beginning with an interstellar ice structure and composition, we track ice destruction due to exposure to UV and heat as dust grains undergo diffusion, settling, and drift within a protoplanetary disk.  We have synthesized the ice survival outcomes from thousands of dust trajectories for different sized particles in different disk regions, and find that ice destruction is generally inefficient except for small grains in the inner few tens of au.

Our modeling supports that the inheritance of pristine interstellar material can indeed explain interstellar-like signatures (e.g. composition and isotopic ratios) measured in comets and primitive meteorites \citep[e.g.][]{Bockelee-Morvan2000, Yurimoto2004, Altwegg2017}.  For our adopted disk physical model, inheritance seems to require a scenario in which icy pebbles form at larger disk radii and then drift into the comet-forming zone ($\sim$5--30 au).  Indeed, small grains in the comet-forming zone rapidly lose their ices, and so local assembly is an unlikely path to inheritance.  Our results therefore lend another piece of evidence to support the importance of pebble drift and pebble accretion to the formation of Solar System bodies \citep[e.g.][]{Lambrechts2012, Lambrechts2014, Levison2015, Johansen2015, Johansen2021}.

The incorporation of interstellar ices into comets and asteroids raises the possibility that molecules synthesized in the ISM can ultimately be delivered to terrestrial planets via impact.  Thus, the wealth of prebiotically interesting molecules detected in pre- and protostellar sources \citep[see e.g.][]{Herbst2009} may indeed be relevant to origins of life chemistry.  Still, better characterization of the photodissociation properties of larger organics are needed to evaluate their survival prospects.  Our models also suggest that small grains at large disk radii may be favorable to a robust in situ ice chemistry due to the abundance of ice photodissociation products and the lack of efficient desorption.  This could complicate interpretations of interstellar isotopic ratios seen in complex organics or refractory organic material in comets and meteorites.  A thorough treatment of ice chemistry accompanying particle dynamics will be the subject of future work.

\acknowledgments 
We thank the anonymous referee for helpful feedback on this manuscript.  We are grateful to Karin \"Oberg for helpful conversations about this project.  J.B.B. acknowledges support from NASA through the NASA Hubble Fellowship grant \#HST-HF2-51429.001-A awarded by the Space Telescope Science Institute, which is operated by the Association of Universities for Research in Astronomy, Incorporated, under NASA contract NAS5-26555. F.J.C. acknowledges support from NASA through the Exoplanetary Research and Emerging World Programs.  The results reported herein benefited from collaborations and/or information exchange within NASA’s Nexus for Exoplanet System Science (NExSS) research coordination network sponsored by NASAs Science Mission Directorate.

\software{
{\fontfamily{qcr}\selectfont Matplotlib} \citep{Hunter2007},
{\fontfamily{qcr}\selectfont NumPy} \citep{VanDerWalt2011},
{\fontfamily{qcr}\selectfont RADMC-3D} \citep{Dullemond2012},
{\fontfamily{qcr}\selectfont Scipy} \citep{SciPy2020}.
}

\FloatBarrier
\clearpage

\appendix 

\section{Fiducial disk physical structure}
\label{sec:app_structure}
\begin{figure}[h!]
\begin{centering}
    \includegraphics[width=0.8\linewidth]{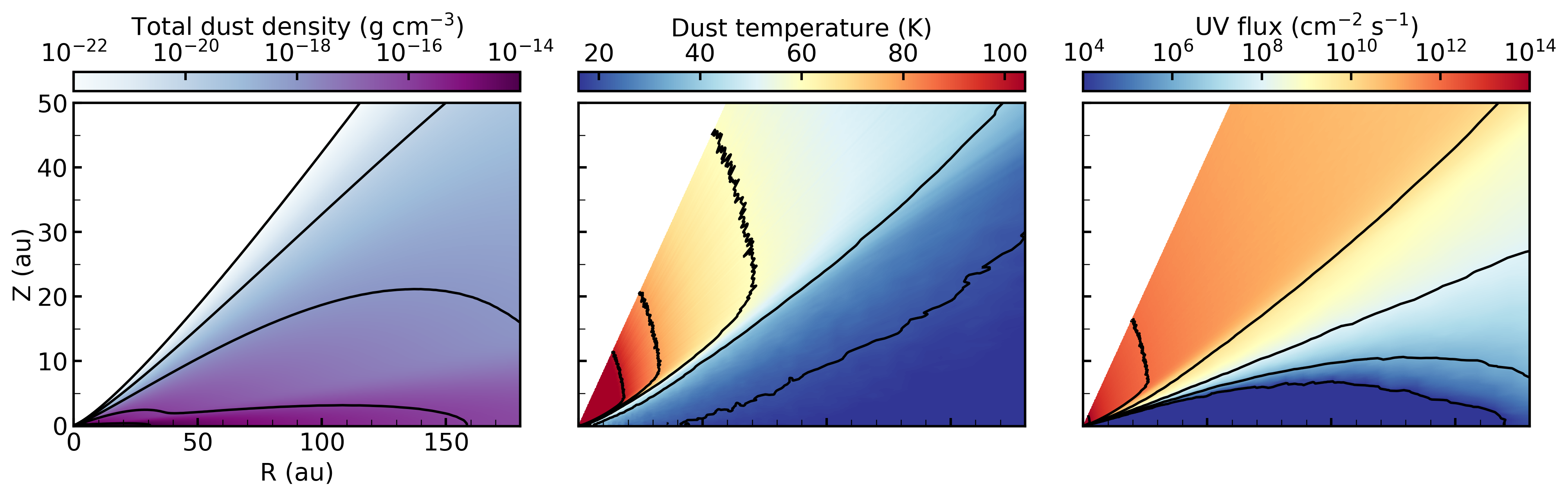}
    \caption{Dust density, dust temperature, and UV flux distributions for the fiducial disk model.}
    \label{fig:disk_structure}
\end{centering}
\end{figure}

Figure \ref{fig:disk_structure} shows the dust density, dust temperature, and UV flux profiles for the disk model described in Section \ref{subsec:model_structure}.

 \FloatBarrier
 
\section{Grain evolution timescales}
\label{sec:app_growth_times}
\begin{figure}
\begin{centering}
    \includegraphics[width=0.4\linewidth]{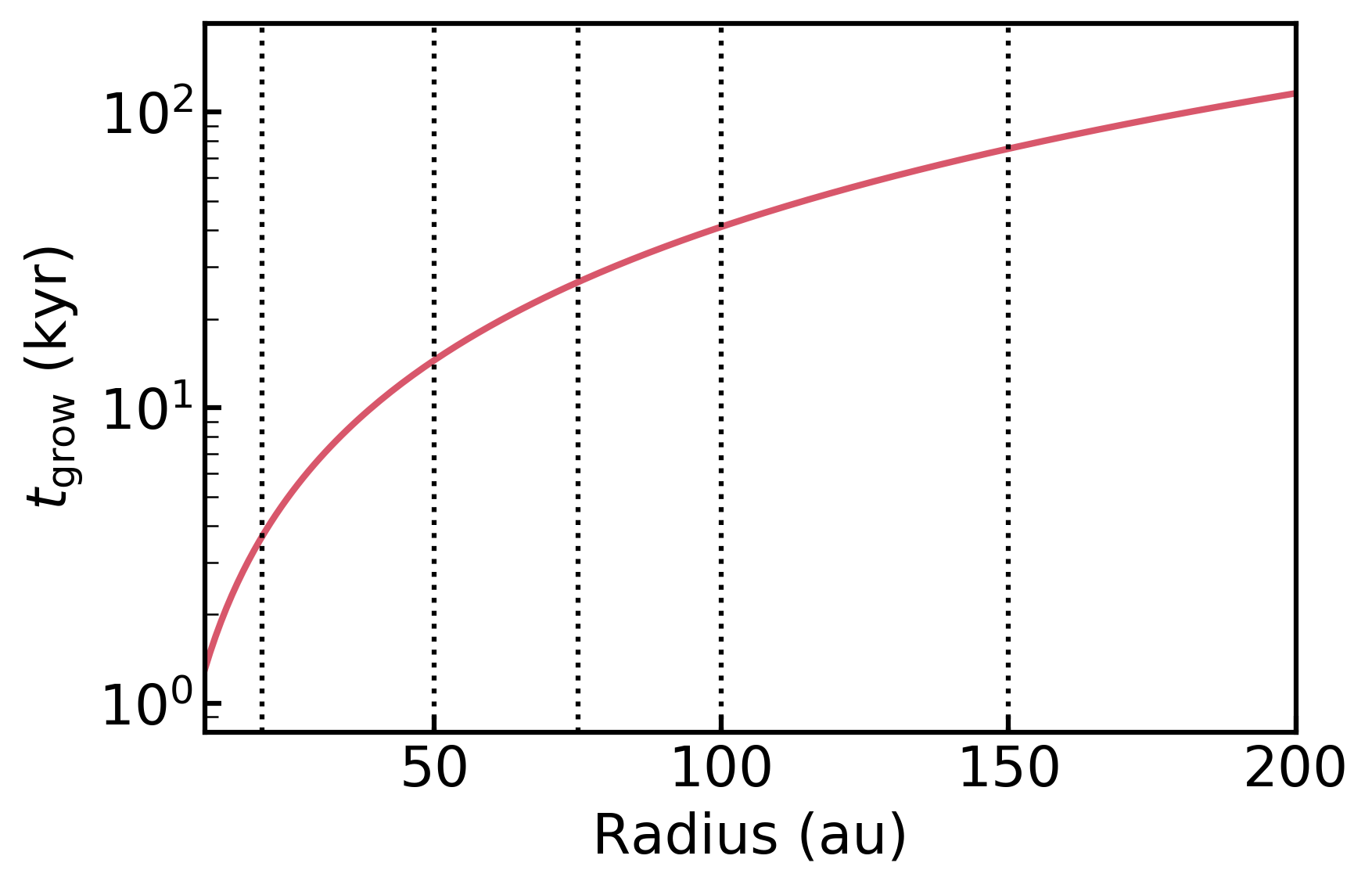}
    \caption{Time required for a grain to grow in size by a factor of 10, as a function of radius in the disk.  Vertical dotted lines indicate the starting radii of particles in our simulations.}
    \label{fig:grain_tau}
\end{centering}
\end{figure}

A particle's trajectory in the disk is strongly dependent on its size.  Therefore, the dynamic trajectories described in Section \ref{subsec:dynamics} are valid only for the amount of time that a particle spends at a given size.  Here we estimate grain growth timescales in order to determine a plausible timescale to track the chemistry for a given grain size.  Following \citet{Birnstiel2012}, the time required for a particle to grow in size by a factor of 10 can be approximated as
\begin{equation}
    t_{\mathrm{grow}} = \tau_{\mathrm{grow}}\mathrm{ln}(10) = \frac{\mathrm{ln}(10)}{\epsilon\sqrt{GM_\star r^3}},
\end{equation}
where $\tau_{\mathrm{grow}}$ is the particle growth e-folding timescale and $\epsilon$ is the dust to gas ratio, assumed to be 10$^{-2}$ here.  Note that the growth timescale is the same for all particle sizes.  Figure \ref{fig:grain_tau} shows the resulting growth timescale as a function of disk radius, which ranges from a few kyr at 20 au to $\sim$75 kyr at 150 au.  

Note, however, that this approximation is derived for particles in the disk midplane.  Grains that undergo excursions to lower-density regions at higher elevations will experience slower growth, impacting especially small grains.  We therefore assume that a timescale of 100 kyr should conservatively encompass the time that a particle spends as a given size. 

\FloatBarrier

\section{Trends with disk evolution}
\label{sec:time_dependence}
\begin{figure*}
\begin{centering}
    \includegraphics[width=\linewidth]{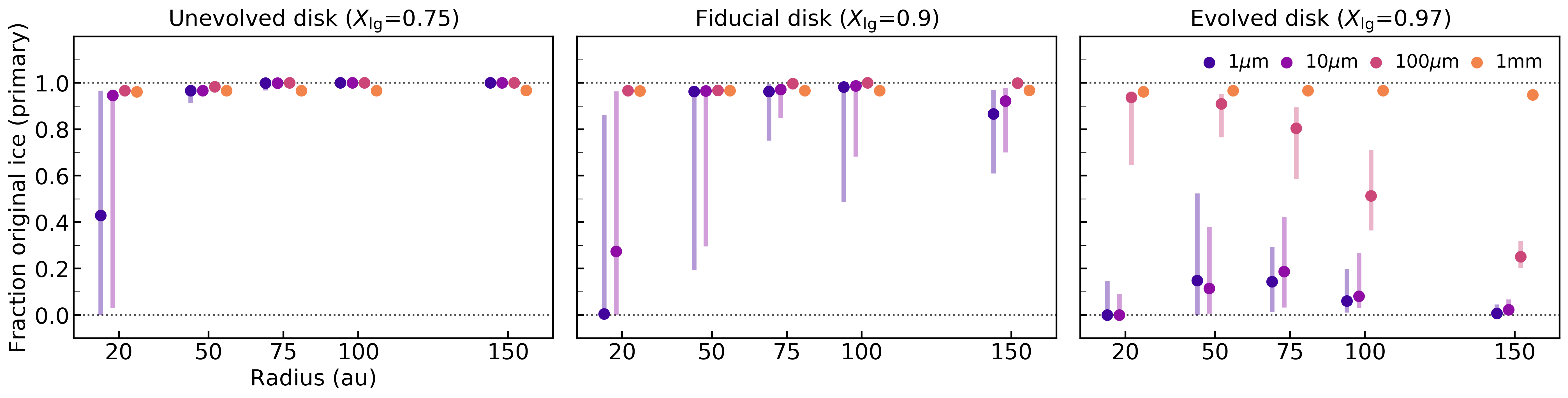}
    \caption{Primary ice fractions remaining at 100 kyr for disk models with varying large-grain fractions $X_\mathrm{lg}$, mimicking less evolved through more evolved disk structures.  For each combination of grain sizes and initial radii, scatter points represent the median survival outcome and shaded bars show the 25-75th percentile range.}
    \label{fig:evol_compare}
\end{centering}
\end{figure*}

The adopted disk structure will impact ice destruction behavior since ice loss is primarily driven by UV exposure, which in turn depends strongly on the distribution of small grains across the disk.  Thus, ice destruction should become more efficient over the disk lifetime as grain growth and settling reduce the small grain population.  Here we explore trends in ice survival for different disk evolutionary stages by varying the fraction of dust partitioned in large grains $X_{\mathrm{lg}}$.  We simulate a less and more evolved disk by adopting an $X_{\mathrm{lg}}$ of 0.75 and 0.97, respectively, compared to the fiducial value of 0.9.  Figure \ref{fig:evol_compare} shows a comparison of the ice survival outcomes for these cases.

In contrast to the fiducial model, which shows ice photodissociation at the tens of percent level in the outer disk, ice destruction is no longer important beyond a few tens of au in the unevolved disk model.  A larger population of small dust results in decreased UV penetration, and so ice photodissociation even in the outer disk is not efficient.  Ice destruction remains important for 1$\mu$m grains at 20 au, though to a lesser extent than in the fiducial model.  Note that while the total survival fraction of primary ices is around 40\%, molecules other than CO and CO$_2$ are generally preserved at just a $\sim$10--20\% level.  Therefore, as in the fiducial model, if a planetesimal body is assembling at 20 au starting from small grains, most pristine ice (apart from CO and CO$_2$) should be lost.  Typical 10$\mu$m grains at 20 au show much less ice loss than in the fiducial model: while some grains can lose almost all their ices, the median particle loses just a few percent of the total ice mantle.  Thus, ice destruction will only be efficient for the smallest grains ($\leq$1$\mu$m) in such an unevolved disk.

In the evolved disk model, ice destruction is much more efficient than in the fiducial model.  1-10$\mu$m grains lose almost all of their ices at all disk radii, and 100$\mu$m grains lose $\gtrsim$50\% of their ice mantle beyond 100 au.  The increase in UV penetration is relatively more important in the less-shielded outer disk compared to the inner disk, resulting in a larger increase in ice destruction at larger radii.  1mm sized grains remain virtually unaffected by ice destruction even with the increased UV penetration of the evolved disk model, indicating that ices incorporated into larger solids in the disk midplane should be quite robust to destruction.

\FloatBarrier
\clearpage
\bibliography{references}

\begin{thebibliography}{}
\expandafter\ifx\csname natexlab\endcsname\relax\def\natexlab#1{#1}\fi
\providecommand{\url}[1]{\href{#1}{#1}}
\providecommand{\dodoi}[1]{doi:~\href{http://doi.org/#1}{\nolinkurl{#1}}}
\providecommand{\doeprint}[1]{\href{http://ascl.net/#1}{\nolinkurl{http://ascl.net/#1}}}
\providecommand{\doarXiv}[1]{\href{https://arxiv.org/abs/#1}{\nolinkurl{https://arxiv.org/abs/#1}}}

\bibitem[{{Aikawa} \& {Herbst}(1999)}]{Aikawa1999}
{Aikawa}, Y., \& {Herbst}, E. 1999, \aap, 351, 233

\bibitem[{{Alexander} {et~al.}(2012){Alexander}, {Bowden}, {Fogel}, {Howard},
  {Herd}, \& {Nittler}}]{Alexander2012}
{Alexander}, C.~M.~O., {Bowden}, R., {Fogel}, M.~L., {et~al.} 2012, Science,
  337, 721, \dodoi{10.1126/science.1223474}

\bibitem[{{Alexander} {et~al.}(2017){Alexander}, {Nittler}, {Davidson}, \&
  {Ciesla}}]{Alexander2017}
{Alexander}, C. M.~O., {Nittler}, L.~R., {Davidson}, J., \& {Ciesla}, F.~J.
  2017, Meteoritics and Planetary Science, 52, 1797, \dodoi{10.1111/maps.12891}

\bibitem[{{Altwegg} {et~al.}(2017){Altwegg}, {Balsiger}, {Berthelier},
  {Bieler}, {Calmonte}, {De Keyser}, {Fiethe}, {Fuselier}, {Gasc}, {Gombosi},
  {Owen}, {Le Roy}, {Rubin}, {S{\'e}mon}, \& {Tzou}}]{Altwegg2017}
{Altwegg}, K., {Balsiger}, H., {Berthelier}, J.~J., {et~al.} 2017,
  Philosophical Transactions of the Royal Society of London Series A, 375,
  20160253, \dodoi{10.1098/rsta.2016.0253}

\bibitem[{{Anders}(1989)}]{Anders1989}
{Anders}, E. 1989, \nat, 342, 255, \dodoi{10.1038/342255a0}

\bibitem[{{Andersson} \& {van Dishoeck}(2008)}]{Andersson2008}
{Andersson}, S., \& {van Dishoeck}, E.~F. 2008, \aap, 491, 907,
  \dodoi{10.1051/0004-6361:200810374}

\bibitem[{{Andrews} {et~al.}(2011){Andrews}, {Wilner}, {Espaillat}, {Hughes},
  {Dullemond}, {McClure}, {Qi}, \& {Brown}}]{Andrews2011}
{Andrews}, S.~M., {Wilner}, D.~J., {Espaillat}, C., {et~al.} 2011, \apj, 732,
  42, \dodoi{10.1088/0004-637X/732/1/42}

\bibitem[{{Andrews} {et~al.}(2012){Andrews}, {Wilner}, {Hughes}, {Qi},
  {Rosenfeld}, {{\"O}berg}, {Birnstiel}, {Espaillat}, {Cieza}, {Williams},
  {Lin}, \& {Ho}}]{Andrews2012}
{Andrews}, S.~M., {Wilner}, D.~J., {Hughes}, A.~M., {et~al.} 2012, \apj, 744,
  162, \dodoi{10.1088/0004-637X/744/2/162}

\bibitem[{{Bergner} {et~al.}(2017){Bergner}, {{\"O}berg}, {Garrod}, \&
  {Graninger}}]{Bergner2017}
{Bergner}, J.~B., {{\"O}berg}, K.~I., {Garrod}, R.~T., \& {Graninger}, D.~M.
  2017, \apj, 841, 120, \dodoi{10.3847/1538-4357/aa72f6}

\bibitem[{{Bertin} {et~al.}(2012){Bertin}, {Fayolle}, {Romanzin}, {{\"O}berg},
  {Michaut}, {Moudens}, {Philippe}, {Jeseck}, {Linnartz}, \&
  {Fillion}}]{Bertin2012}
{Bertin}, M., {Fayolle}, E.~C., {Romanzin}, C., {et~al.} 2012, Physical
  Chemistry Chemical Physics (Incorporating Faraday Transactions), 14, 9929,
  \dodoi{10.1039/C2CP41177F}

\bibitem[{{Bethell} \& {Bergin}(2011)}]{Bethell2011}
{Bethell}, T.~J., \& {Bergin}, E.~A. 2011, \apj, 739, 78,
  \dodoi{10.1088/0004-637X/739/2/78}

\bibitem[{{Birnstiel} {et~al.}(2012){Birnstiel}, {Klahr}, \&
  {Ercolano}}]{Birnstiel2012}
{Birnstiel}, T., {Klahr}, H., \& {Ercolano}, B. 2012, \aap, 539, A148,
  \dodoi{10.1051/0004-6361/201118136}

\bibitem[{{Bockel{\'e}e-Morvan} {et~al.}(2004){Bockel{\'e}e-Morvan},
  {Crovisier}, {Mumma}, \& {Weaver}}]{Bockelee-Morvan2004}
{Bockel{\'e}e-Morvan}, D., {Crovisier}, J., {Mumma}, M.~J., \& {Weaver}, H.~A.
  2004, {The composition of cometary volatiles}, ed. M.~C. {Festou}, H.~U.
  {Keller}, \& H.~A. {Weaver}, 391

\bibitem[{{Bockel{\'e}e-Morvan} {et~al.}(2002){Bockel{\'e}e-Morvan}, {Gautier},
  {Hersant}, {Hur{\'e}}, \& {Robert}}]{Bockelee-Morvan2002}
{Bockel{\'e}e-Morvan}, D., {Gautier}, D., {Hersant}, F., {Hur{\'e}}, J.~M., \&
  {Robert}, F. 2002, \aap, 384, 1107, \dodoi{10.1051/0004-6361:20020086}

\bibitem[{{Bockel{\'e}e-Morvan} {et~al.}(2000){Bockel{\'e}e-Morvan}, {Lis},
  {Wink}, {Despois}, {Crovisier}, {Bachiller}, {Benford}, {Biver}, {Colom},
  {Davies}, {G{\'e}rard}, {Germain}, {Houde}, {Mehringer}, {Moreno}, {Paubert},
  {Phillips}, \& {Rauer}}]{Bockelee-Morvan2000}
{Bockel{\'e}e-Morvan}, D., {Lis}, D.~C., {Wink}, J.~E., {et~al.} 2000, \aap,
  353, 1101

\bibitem[{{Bockel{\'e}e-Morvan} {et~al.}(2015){Bockel{\'e}e-Morvan},
  {Calmonte}, {Charnley}, {Duprat}, {Engrand}, {Gicquel}, {H{\"a}ssig},
  {Jehin}, {Kawakita}, {Marty}, {Milam}, {Morse}, {Rousselot}, {Sheridan}, \&
  {Wirstr{\"o}m}}]{Bockelee-Morvan2015}
{Bockel{\'e}e-Morvan}, D., {Calmonte}, U., {Charnley}, S., {et~al.} 2015, \ssr,
  197, 47, \dodoi{10.1007/s11214-015-0156-9}

\bibitem[{{Brownlee} {et~al.}(2006){Brownlee}, {Tsou}, {Al{\'e}on},
  {Alexander}, {Araki}, {Bajt}, {Baratta}, {Bastien}, {Bland}, {Bleuet},
  {Borg}, {Bradley}, {Brearley}, {Brenker}, {Brennan}, {Bridges}, {Browning},
  {Brucato}, {Bullock}, {Burchell}, {Busemann}, {Butterworth}, {Chaussidon},
  {Cheuvront}, {Chi}, {Cintala}, {Clark}, {Clemett}, {Cody}, {Colangeli},
  {Cooper}, {Cordier}, {Daghlian}, {Dai}, {D'Hendecourt}, {Djouadi},
  {Dominguez}, {Duxbury}, {Dworkin}, {Ebel}, {Economou}, {Fakra}, {Fairey},
  {Fallon}, {Ferrini}, {Ferroir}, {Fleckenstein}, {Floss}, {Flynn}, {Franchi},
  {Fries}, {Gainsforth}, {Gallien}, {Genge}, {Gilles}, {Gillet}, {Gilmour},
  {Glavin}, {Gounelle}, {Grady}, {Graham}, {Grant}, {Green}, {Grossemy},
  {Grossman}, {Grossman}, {Guan}, {Hagiya}, {Harvey}, {Heck}, {Herzog},
  {Hoppe}, {H{\"o}rz}, {Huth}, {Hutcheon}, {Ignatyev}, {Ishii}, {Ito}, {Jacob},
  {Jacobsen}, {Jacobsen}, {Jones}, {Joswiak}, {Jurewicz}, {Kearsley}, {Keller},
  {Khodja}, {Kilcoyne}, {Kissel}, {Krot}, {Langenhorst}, {Lanzirotti}, {Le},
  {Leshin}, {Leitner}, {Lemelle}, {Leroux}, {Liu}, {Luening}, {Lyon},
  {MacPherson}, {Marcus}, {Marhas}, {Marty}, {Matrajt}, {McKeegan}, {Meibom},
  {Mennella}, {Messenger}, {Messenger}, {Mikouchi}, {Mostefaoui}, {Nakamura},
  {Nakano}, {Newville}, {Nittler}, {Ohnishi}, {Ohsumi}, {Okudaira},
  {Papanastassiou}, {Palma}, {Palumbo}, {Pepin}, {Perkins}, {Perronnet},
  {Pianetta}, {Rao}, {Rietmeijer}, {Robert}, {Rost}, {Rotundi}, {Ryan},
  {Sandford}, {Schwandt}, {See}, {Schlutter}, {Sheffield-Parker},
  {Simionovici}, {Simon}, {Sitnitsky}, {Snead}, {Spencer}, {Stadermann},
  {Steele}, {Stephan}, {Stroud}, {Susini}, {Sutton}, {Suzuki}, {Taheri},
  {Taylor}, {Teslich}, {Tomeoka}, {Tomioka}, {Toppani}, {Trigo-Rodr{\'\i}guez},
  {Troadec}, {Tsuchiyama}, {Tuzzolino}, {Tyliszczak}, {Uesugi}, {Velbel},
  {Vellenga}, {Vicenzi}, {Vincze}, {Warren}, {Weber}, {Weisberg}, {Westphal},
  {Wirick}, {Wooden}, {Wopenka}, {Wozniakiewicz}, {Wright}, {Yabuta}, {Yano},
  {Young}, {Zare}, {Zega}, {Ziegler}, {Zimmerman}, {Zinner}, \&
  {Zolensky}}]{Brownlee2006}
{Brownlee}, D., {Tsou}, P., {Al{\'e}on}, J., {et~al.} 2006, Science, 314, 1711,
  \dodoi{10.1126/science.1135840}

\bibitem[{{Busemann} {et~al.}(2006){Busemann}, {Young}, {O'D. Alexander},
  {Hoppe}, {Mukhopadhyay}, \& {Nittler}}]{Busemann2006}
{Busemann}, H., {Young}, A.~F., {O'D. Alexander}, C.~M., {et~al.} 2006,
  Science, 312, 727, \dodoi{10.1126/science.1123878}

\bibitem[{{Ciesla}(2010)}]{Ciesla2010}
{Ciesla}, F.~J. 2010, \apj, 723, 514, \dodoi{10.1088/0004-637X/723/1/514}

\bibitem[{{Ciesla}(2011)}]{Ciesla2011}
---. 2011, \apj, 740, 9, \dodoi{10.1088/0004-637X/740/1/9}

\bibitem[{{Ciesla} \& {Sandford}(2012)}]{Ciesla2012}
{Ciesla}, F.~J., \& {Sandford}, S.~A. 2012, Science, 336, 452,
  \dodoi{10.1126/science.1217291}

\bibitem[{{Cleeves} {et~al.}(2013){Cleeves}, {Adams}, \&
  {Bergin}}]{Cleeves2013}
{Cleeves}, L.~I., {Adams}, F.~C., \& {Bergin}, E.~A. 2013, \apj, 772, 5,
  \dodoi{10.1088/0004-637X/772/1/5}

\bibitem[{{Cleeves} {et~al.}(2014){Cleeves}, {Bergin}, {Alexander}, {Du},
  {Graninger}, {{\"O}berg}, \& {Harries}}]{Cleeves2014}
{Cleeves}, L.~I., {Bergin}, E.~A., {Alexander}, C. M.~O.~D., {et~al.} 2014,
  Science, 345, 1590, \dodoi{10.1126/science.1258055}

\bibitem[{{Cleeves} {et~al.}(2015){Cleeves}, {Bergin}, {Qi}, {Adams}, \&
  {{\"O}berg}}]{Cleeves2015}
{Cleeves}, L.~I., {Bergin}, E.~A., {Qi}, C., {Adams}, F.~C., \& {{\"O}berg},
  K.~I. 2015, \apj, 799, 204, \dodoi{10.1088/0004-637X/799/2/204}

\bibitem[{{Dartois} {et~al.}(2003){Dartois}, {Dutrey}, \&
  {Guilloteau}}]{Dartois2003}
{Dartois}, E., {Dutrey}, A., \& {Guilloteau}, S. 2003, \aap, 399, 773,
  \dodoi{10.1051/0004-6361:20021638}

\bibitem[{{Davidsson} {et~al.}(2016){Davidsson}, {Sierks}, {G{\"u}ttler},
  {Marzari}, {Pajola}, {Rickman}, {A'Hearn}, {Auger}, {El-Maarry}, {Fornasier},
  {Guti{\'e}rrez}, {Keller}, {Massironi}, {Snodgrass}, {Vincent}, {Barbieri},
  {Lamy}, {Rodrigo}, {Koschny}, {Barucci}, {Bertaux}, {Bertini}, {Cremonese},
  {Da Deppo}, {Debei}, {De Cecco}, {Feller}, {Fulle}, {Groussin}, {Hviid},
  {H{\"o}fner}, {Ip}, {Jorda}, {Knollenberg}, {Kovacs}, {Kramm}, {K{\"u}hrt},
  {K{\"u}ppers}, {La Forgia}, {Lara}, {Lazzarin}, {Lopez Moreno},
  {Moissl-Fraund}, {Mottola}, {Naletto}, {Oklay}, {Thomas}, \&
  {Tubiana}}]{Davidsson2016}
{Davidsson}, B.~J.~R., {Sierks}, H., {G{\"u}ttler}, C., {et~al.} 2016, \aap,
  592, A63, \dodoi{10.1051/0004-6361/201526968}

\bibitem[{{Drozdovskaya} {et~al.}(2019){Drozdovskaya}, {van Dishoeck}, {Rubin},
  {J{\o}rgensen}, \& {Altwegg}}]{Drozdovskaya2019}
{Drozdovskaya}, M.~N., {van Dishoeck}, E.~F., {Rubin}, M., {J{\o}rgensen},
  J.~K., \& {Altwegg}, K. 2019, \mnras, 490, 50, \dodoi{10.1093/mnras/stz2430}

\bibitem[{{Drozdovskaya} {et~al.}(2016){Drozdovskaya}, {Walsh}, {van Dishoeck},
  {Furuya}, {Marboeuf}, {Thiabaud}, {Harsono}, \& {Visser}}]{Drozdovskaya2016}
{Drozdovskaya}, M.~N., {Walsh}, C., {van Dishoeck}, E.~F., {et~al.} 2016,
  \mnras, 462, 977, \dodoi{10.1093/mnras/stw1632}

\bibitem[{{Drozdovskaya} {et~al.}(2014){Drozdovskaya}, {Walsh}, {Visser},
  {Harsono}, \& {van Dishoeck}}]{Drozdovskaya2014}
{Drozdovskaya}, M.~N., {Walsh}, C., {Visser}, R., {Harsono}, D., \& {van
  Dishoeck}, E.~F. 2014, \mnras, 445, 913, \dodoi{10.1093/mnras/stu1789}

\bibitem[{{Drozdovskaya} {et~al.}(2021){Drozdovskaya}, {Schroeder I}, {Rubin},
  {Altwegg}, {van Dishoeck}, {Kulterer}, {De Keyser}, {Fuselier}, \&
  {Combi}}]{Drozdovskaya2021}
{Drozdovskaya}, M.~N., {Schroeder I}, I. R.~H.~G., {Rubin}, M., {et~al.} 2021,
  \mnras, 500, 4901, \dodoi{10.1093/mnras/staa3387}

\bibitem[{{Dullemond} \& {Dominik}(2004)}]{Dullemond2004}
{Dullemond}, C.~P., \& {Dominik}, C. 2004, \aap, 421, 1075,
  \dodoi{10.1051/0004-6361:20040284}

\bibitem[{{Dullemond} {et~al.}(2012){Dullemond}, {Juhasz}, {Pohl}, {Sereshti},
  {Shetty}, {Peters}, {Commercon}, \& {Flock}}]{Dullemond2012}
{Dullemond}, C.~P., {Juhasz}, A., {Pohl}, A., {et~al.} 2012, {RADMC-3D: A
  multi-purpose radiative transfer tool}.
\newblock \doeprint{1202.015}

\bibitem[{{Dutrey} {et~al.}(1997){Dutrey}, {Guilloteau}, \&
  {Guelin}}]{Dutrey1997}
{Dutrey}, A., {Guilloteau}, S., \& {Guelin}, M. 1997, \aap, 317, L55

\bibitem[{{Fayolle} {et~al.}(2011){Fayolle}, {{\"O}berg}, {Cuppen}, {Visser},
  \& {Linnartz}}]{Fayolle2011}
{Fayolle}, E.~C., {{\"O}berg}, K.~I., {Cuppen}, H.~M., {Visser}, R., \&
  {Linnartz}, H. 2011, \aap, 529, A74, \dodoi{10.1051/0004-6361/201016121}

\bibitem[{{Furuya} \& {Aikawa}(2018)}]{Furuya2018}
{Furuya}, K., \& {Aikawa}, Y. 2018, \apj, 857, 105,
  \dodoi{10.3847/1538-4357/aab768}

\bibitem[{{Garrod}(2013)}]{Garrod2013}
{Garrod}, R.~T. 2013, \apj, 765, 60, \dodoi{10.1088/0004-637X/765/1/60}

\bibitem[{{Garrod}(2019)}]{Garrod2019}
---. 2019, \apj, 884, 69, \dodoi{10.3847/1538-4357/ab418e}

\bibitem[{{Habing}(1968)}]{Habing1968}
{Habing}, H.~J. 1968, \bain, 19, 421

\bibitem[{{Hartmann} {et~al.}(1998){Hartmann}, {Calvet}, {Gullbring}, \&
  {D'Alessio}}]{Hartmann1998}
{Hartmann}, L., {Calvet}, N., {Gullbring}, E., \& {D'Alessio}, P. 1998, \apj,
  495, 385, \dodoi{10.1086/305277}

\bibitem[{{Hasegawa} {et~al.}(1992){Hasegawa}, {Herbst}, \&
  {Leung}}]{Hasegawa1992}
{Hasegawa}, T.~I., {Herbst}, E., \& {Leung}, C.~M. 1992, \apjs, 82, 167,
  \dodoi{10.1086/191713}

\bibitem[{{Heays} {et~al.}(2017){Heays}, {Bosman}, \& {van
  Dishoeck}}]{Heays2017}
{Heays}, A.~N., {Bosman}, A.~D., \& {van Dishoeck}, E.~F. 2017, \aap, 602,
  A105, \dodoi{10.1051/0004-6361/201628742}

\bibitem[{{Herbst} \& {van Dishoeck}(2009)}]{Herbst2009}
{Herbst}, E., \& {van Dishoeck}, E.~F. 2009, \araa, 47, 427,
  \dodoi{10.1146/annurev-astro-082708-101654}

\bibitem[{{Herczeg} {et~al.}(2002){Herczeg}, {Linsky}, {Valenti},
  {Johns-Krull}, \& {Wood}}]{Herczeg2002}
{Herczeg}, G.~J., {Linsky}, J.~L., {Valenti}, J.~A., {Johns-Krull}, C.~M., \&
  {Wood}, B.~E. 2002, \apj, 572, 310, \dodoi{10.1086/339731}

\bibitem[{{Herczeg} {et~al.}(2004){Herczeg}, {Wood}, {Linsky}, {Valenti}, \&
  {Johns-Krull}}]{Herczeg2004}
{Herczeg}, G.~J., {Wood}, B.~E., {Linsky}, J.~L., {Valenti}, J.~A., \&
  {Johns-Krull}, C.~M. 2004, \apj, 607, 369, \dodoi{10.1086/383340}

\bibitem[{{Hincelin} {et~al.}(2013){Hincelin}, {Wakelam}, {Commer{\c{c}}on},
  {Hersant}, \& {Guilloteau}}]{Hincelin2013}
{Hincelin}, U., {Wakelam}, V., {Commer{\c{c}}on}, B., {Hersant}, F., \&
  {Guilloteau}, S. 2013, \apj, 775, 44, \dodoi{10.1088/0004-637X/775/1/44}

\bibitem[{{Hollis} {et~al.}(2000){Hollis}, {Lovas}, \& {Jewell}}]{Hollis2000}
{Hollis}, J.~M., {Lovas}, F.~J., \& {Jewell}, P.~R. 2000, \apjl, 540, L107,
  \dodoi{10.1086/312881}

\bibitem[{{Huang} {et~al.}(2018){Huang}, {Andrews}, {Cleeves}, {{\"O}berg},
  {Wilner}, {Bai}, {Birnstiel}, {Carpenter}, {Hughes}, {Isella}, {P{\'e}rez},
  {Ricci}, \& {Zhu}}]{Huang2018}
{Huang}, J., {Andrews}, S.~M., {Cleeves}, L.~I., {et~al.} 2018, \apj, 852, 122,
  \dodoi{10.3847/1538-4357/aaa1e7}

\bibitem[{{Hunter}(2007)}]{Hunter2007}
{Hunter}, J.~D. 2007, Computing in Science and Engineering, 9, 90,
  \dodoi{10.1109/MCSE.2007.55}

\bibitem[{{Johansen} {et~al.}(2015){Johansen}, {Mac Low}, {Lacerda}, \&
  {Bizzarro}}]{Johansen2015}
{Johansen}, A., {Mac Low}, M.-M., {Lacerda}, P., \& {Bizzarro}, M. 2015,
  Science Advances, 1, 1500109, \dodoi{10.1126/sciadv.1500109}

\bibitem[{{Johansen} {et~al.}(2021){Johansen}, {Ronnet}, {Bizzarro},
  {Schiller}, {Lambrechts}, {Nordlund}, \& {Lammer}}]{Johansen2021}
{Johansen}, A., {Ronnet}, T., {Bizzarro}, M., {et~al.} 2021, arXiv e-prints,
  arXiv:2102.08611.
\newblock \doarXiv{2102.08611}

\bibitem[{{J{\o}rgensen} {et~al.}(2012){J{\o}rgensen}, {Favre}, {Bisschop},
  {Bourke}, {van Dishoeck}, \& {Schmalzl}}]{Jorgensen2012}
{J{\o}rgensen}, J.~K., {Favre}, C., {Bisschop}, S.~E., {et~al.} 2012, \apjl,
  757, L4, \dodoi{10.1088/2041-8205/757/1/L4}

\bibitem[{{Kahane} {et~al.}(2013){Kahane}, {Ceccarelli}, {Faure}, \&
  {Caux}}]{Kahane2013}
{Kahane}, C., {Ceccarelli}, C., {Faure}, A., \& {Caux}, E. 2013, \apjl, 763,
  L38, \dodoi{10.1088/2041-8205/763/2/L38}

\bibitem[{Krot {et~al.}(2020)Krot, Nagashima, Lyons, Lee, \&
  Bizzarro}]{Krot2020}
Krot, A.~N., Nagashima, K., Lyons, J.~R., Lee, J.-E., \& Bizzarro, M. 2020,
  Science Advances, 6, \dodoi{10.1126/sciadv.aay2724}

\bibitem[{{Lambrechts} \& {Johansen}(2012)}]{Lambrechts2012}
{Lambrechts}, M., \& {Johansen}, A. 2012, \aap, 544, A32,
  \dodoi{10.1051/0004-6361/201219127}

\bibitem[{{Lambrechts} \& {Johansen}(2014)}]{Lambrechts2014}
---. 2014, \aap, 572, A107, \dodoi{10.1051/0004-6361/201424343}

\bibitem[{{Lambrechts} {et~al.}(2019){Lambrechts}, {Morbidelli}, {Jacobson},
  {Johansen}, {Bitsch}, {Izidoro}, \& {Raymond}}]{Lambrechts2019}
{Lambrechts}, M., {Morbidelli}, A., {Jacobson}, S.~A., {et~al.} 2019, \aap,
  627, A83, \dodoi{10.1051/0004-6361/201834229}

\bibitem[{{Lee} {et~al.}(2008){Lee}, {Bergin}, \& {Lyons}}]{Lee2008}
{Lee}, J.-E., {Bergin}, E.~A., \& {Lyons}, J.~R. 2008, Meteoritics and
  Planetary Science, 43, 1351, \dodoi{10.1111/j.1945-5100.2008.tb00702.x}

\bibitem[{{Levison} {et~al.}(2015){Levison}, {Kretke}, {Walsh}, \&
  {Bottke}}]{Levison2015}
{Levison}, H.~F., {Kretke}, K.~A., {Walsh}, K.~J., \& {Bottke}, W.~F. 2015,
  Proceedings of the National Academy of Science, 112, 14180,
  \dodoi{10.1073/pnas.1513364112}

\bibitem[{{Lorek} {et~al.}(2018){Lorek}, {Lacerda}, \& {Blum}}]{Lorek2018}
{Lorek}, S., {Lacerda}, P., \& {Blum}, J. 2018, \aap, 611, A18,
  \dodoi{10.1051/0004-6361/201630175}

\bibitem[{{Lynden-Bell} \& {Pringle}(1974)}]{Lynden-Bell1974}
{Lynden-Bell}, D., \& {Pringle}, J.~E. 1974, \mnras, 168, 603,
  \dodoi{10.1093/mnras/168.3.603}

\bibitem[{{McElroy} {et~al.}(2013){McElroy}, {Walsh}, {Markwick}, {Cordiner},
  {Smith}, \& {Millar}}]{McElroy2013}
{McElroy}, D., {Walsh}, C., {Markwick}, A.~J., {et~al.} 2013, \aap, 550, A36,
  \dodoi{10.1051/0004-6361/201220465}

\bibitem[{{McKeegan} {et~al.}(2006){McKeegan}, {Al{\'e}on}, {Bradley},
  {Brownlee}, {Busemann}, {Butterworth}, {Chaussidon}, {Fallon}, {Floss},
  {Gilmour}, {Gounelle}, {Graham}, {Guan}, {Heck}, {Hoppe}, {Hutcheon}, {Huth},
  {Ishii}, {Ito}, {Jacobsen}, {Kearsley}, {Leshin}, {Liu}, {Lyon}, {Marhas},
  {Marty}, {Matrajt}, {Meibom}, {Messenger}, {Mostefaoui}, {Mukhopadhyay},
  {Nakamura-Messenger}, {Nittler}, {Palma}, {Pepin}, {Papanastassiou},
  {Robert}, {Schlutter}, {Snead}, {Stadermann}, {Stroud}, {Tsou}, {Westphal},
  {Young}, {Ziegler}, {Zimmermann}, \& {Zinner}}]{McKeegan2006}
{McKeegan}, K.~D., {Al{\'e}on}, J., {Bradley}, J., {et~al.} 2006, Science, 314,
  1724, \dodoi{10.1126/science.1135992}

\bibitem[{{Mu{\~n}oz Caro} {et~al.}(2010){Mu{\~n}oz Caro},
  {Jim{\'e}nez-Escobar}, {Mart{\'\i}n-Gago}, {Rogero}, {Atienza}, {Puertas},
  {Sobrado}, \& {Torres-Redondo}}]{Munoz-Caro2010}
{Mu{\~n}oz Caro}, G.~M., {Jim{\'e}nez-Escobar}, A., {Mart{\'\i}n-Gago},
  J.~{\'A}., {et~al.} 2010, \aap, 522, A108,
  \dodoi{10.1051/0004-6361/200912462}

\bibitem[{{Noble} {et~al.}(2013){Noble}, {Theule}, {Borget}, {Danger},
  {Chomat}, {Duvernay}, {Mispelaer}, \& {Chiavassa}}]{Noble2013}
{Noble}, J.~A., {Theule}, P., {Borget}, F., {et~al.} 2013, \mnras, 428, 3262,
  \dodoi{10.1093/mnras/sts272}

\bibitem[{{Noble} {et~al.}(2012){Noble}, {Theule}, {Mispelaer}, {Duvernay},
  {Danger}, {Congiu}, {Dulieu}, \& {Chiavassa}}]{Noble2012}
{Noble}, J.~A., {Theule}, P., {Mispelaer}, F., {et~al.} 2012, \aap, 543, A5,
  \dodoi{10.1051/0004-6361/201219437}

\bibitem[{{{\"O}berg} {et~al.}(2011){{\"O}berg}, {Boogert}, {Pontoppidan}, {van
  den Broek}, {van Dishoeck}, {Bottinelli}, {Blake}, \& {Evans}}]{Oberg2011}
{{\"O}berg}, K.~I., {Boogert}, A.~C.~A., {Pontoppidan}, K.~M., {et~al.} 2011,
  \apj, 740, 109, \dodoi{10.1088/0004-637X/740/2/109}

\bibitem[{{Or{\'o}}(1961)}]{Oro1961}
{Or{\'o}}, J. 1961, \nat, 190, 389, \dodoi{10.1038/190389a0}

\bibitem[{{Powner} {et~al.}(2009){Powner}, {Gerland}, \&
  {Sutherland}}]{Powner2009}
{Powner}, M.~W., {Gerland}, B., \& {Sutherland}, J.~D. 2009, \nat, 459, 239,
  \dodoi{10.1038/nature08013}

\bibitem[{{Rosenfeld} {et~al.}(2013){Rosenfeld}, {Andrews}, {Wilner},
  {Kastner}, \& {McClure}}]{Rosenfeld2013}
{Rosenfeld}, K.~A., {Andrews}, S.~M., {Wilner}, D.~J., {Kastner}, J.~H., \&
  {McClure}, M.~K. 2013, \apj, 775, 136, \dodoi{10.1088/0004-637X/775/2/136}

\bibitem[{{Ruaud} \& {Gorti}(2019)}]{Ruaud2019}
{Ruaud}, M., \& {Gorti}, U. 2019, \apj, 885, 146,
  \dodoi{10.3847/1538-4357/ab4996}

\bibitem[{{Ruaud} {et~al.}(2016){Ruaud}, {Wakelam}, \& {Hersant}}]{Ruaud2016}
{Ruaud}, M., {Wakelam}, V., \& {Hersant}, F. 2016, \mnras, 459, 3756,
  \dodoi{10.1093/mnras/stw887}

\bibitem[{{Rubin} {et~al.}(2019){Rubin}, {Bekaert}, {Broadley}, {Drozdovskaya},
  \& {Wampfler}}]{Rubin2019}
{Rubin}, M., {Bekaert}, D.~V., {Broadley}, M.~W., {Drozdovskaya}, M.~N., \&
  {Wampfler}, S.~F. 2019, ACS Earth and Space Chemistry, 3, 1792,
  \dodoi{10.1021/acsearthspacechem.9b00096}

\bibitem[{{Rubin} {et~al.}(2020){Rubin}, {Engrand}, {Snodgrass}, {Weissman},
  {Altwegg}, {Busemann}, {Morbidelli}, \& {Mumma}}]{Rubin2020}
{Rubin}, M., {Engrand}, C., {Snodgrass}, C., {et~al.} 2020, \ssr, 216, 102,
  \dodoi{10.1007/s11214-020-00718-2}

\bibitem[{{Sandford} \& {Allamandola}(1993)}]{Sandford1993}
{Sandford}, S.~A., \& {Allamandola}, L.~J. 1993, \apj, 417, 815,
  \dodoi{10.1086/173362}

\bibitem[{{Schindhelm} {et~al.}(2012){Schindhelm}, {France}, {Herczeg},
  {Bergin}, {Yang}, {Brown}, {Brown}, {Linsky}, \& {Valenti}}]{Schindhelm2012}
{Schindhelm}, E., {France}, K., {Herczeg}, G.~J., {et~al.} 2012, \apjl, 756,
  L23, \dodoi{10.1088/2041-8205/756/1/L23}

\bibitem[{{Semenov} \& {Wiebe}(2011)}]{Semenov2011}
{Semenov}, D., \& {Wiebe}, D. 2011, \apjs, 196, 25,
  \dodoi{10.1088/0067-0049/196/2/25}

\bibitem[{{Simon} {et~al.}(2019){Simon}, {{\"O}berg}, {Rajappan}, \&
  {Maksiutenko}}]{Simon2019}
{Simon}, A., {{\"O}berg}, K.~I., {Rajappan}, M., \& {Maksiutenko}, P. 2019,
  \apj, 883, 21, \dodoi{10.3847/1538-4357/ab32e5}

\bibitem[{{Snyder} \& {Buhl}(1971)}]{Snyder1971}
{Snyder}, L.~E., \& {Buhl}, D. 1971, \apjl, 163, L47, \dodoi{10.1086/180664}

\bibitem[{{Taquet} {et~al.}(2014){Taquet}, {Charnley}, \&
  {Sipil{\"a}}}]{Taquet2014}
{Taquet}, V., {Charnley}, S.~B., \& {Sipil{\"a}}, O. 2014, \apj, 791, 1,
  \dodoi{10.1088/0004-637X/791/1/1}

\bibitem[{{Testi} {et~al.}(2014){Testi}, {Birnstiel}, {Ricci}, {Andrews},
  {Blum}, {Carpenter}, {Dominik}, {Isella}, {Natta}, {Williams}, \&
  {Wilner}}]{Testi2014}
{Testi}, L., {Birnstiel}, T., {Ricci}, L., {et~al.} 2014, in Protostars and
  Planets VI, ed. H.~{Beuther}, R.~S. {Klessen}, C.~P. {Dullemond}, \&
  T.~{Henning}, 339, \dodoi{10.2458/azu_uapress_9780816531240-ch015}

\bibitem[{{Turner}(1989)}]{Turner1989}
{Turner}, B.~E. 1989, \apjs, 70, 539, \dodoi{10.1086/191348}

\bibitem[{{van der Walt} {et~al.}(2011){van der Walt}, {Colbert}, \&
  {Varoquaux}}]{VanDerWalt2011}
{van der Walt}, S., {Colbert}, S.~C., \& {Varoquaux}, G. 2011, Computing in
  Science and Engineering, 13, 22, \dodoi{10.1109/MCSE.2011.37}

\bibitem[{{van Dishoeck} {et~al.}(2014){van Dishoeck}, {Bergin}, {Lis}, \&
  {Lunine}}]{vanDishoeck2014}
{van Dishoeck}, E.~F., {Bergin}, E.~A., {Lis}, D.~C., \& {Lunine}, J.~I. 2014,
  in Protostars and Planets VI, ed. H.~{Beuther}, R.~S. {Klessen}, C.~P.
  {Dullemond}, \& T.~{Henning}, 835,
  \dodoi{10.2458/azu_uapress_9780816531240-ch036}

\bibitem[{{van Dishoeck} {et~al.}(1995){van Dishoeck}, {Blake}, {Jansen}, \&
  {Groesbeck}}]{vanDishoeck1995}
{van Dishoeck}, E.~F., {Blake}, G.~A., {Jansen}, D.~J., \& {Groesbeck}, T.~D.
  1995, \apj, 447, 760, \dodoi{10.1086/175915}

\bibitem[{{Virtanen} {et~al.}(2020){Virtanen}, {Gommers}, {Oliphant},
  {Haberland}, {Reddy}, {Cournapeau}, {Burovski}, {Peterson}, {Weckesser},
  {Bright}, {van der Walt}, {Brett}, {Wilson}, {Jarrod Millman}, {Mayorov},
  {Nelson}, {Jones}, {Kern}, {Larson}, {Carey}, {Polat}, {Feng}, {Moore}, {Vand
  erPlas}, {Laxalde}, {Perktold}, {Cimrman}, {Henriksen}, {Quintero}, {Harris},
  {Archibald}, {Ribeiro}, {Pedregosa}, {van Mulbregt}, \&
  {Contributors}}]{SciPy2020}
{Virtanen}, P., {Gommers}, R., {Oliphant}, T.~E., {et~al.} 2020, Nature
  Methods, 17, 261, \dodoi{https://doi.org/10.1038/s41592-019-0686-2}

\bibitem[{{Visser} {et~al.}(2011){Visser}, {Doty}, \& {van
  Dishoeck}}]{Visser2011}
{Visser}, R., {Doty}, S.~D., \& {van Dishoeck}, E.~F. 2011, \aap, 534, A132,
  \dodoi{10.1051/0004-6361/201117249}

\bibitem[{{Visser} {et~al.}(2009){Visser}, {van Dishoeck}, {Doty}, \&
  {Dullemond}}]{Visser2009}
{Visser}, R., {van Dishoeck}, E.~F., {Doty}, S.~D., \& {Dullemond}, C.~P. 2009,
  \aap, 495, 881, \dodoi{10.1051/0004-6361/200810846}

\bibitem[{{Weidenschilling}(1997)}]{Weidenschilling1997}
{Weidenschilling}, S.~J. 1997, \icarus, 127, 290,
  \dodoi{10.1006/icar.1997.5712}

\bibitem[{{Woitke} {et~al.}(2016){Woitke}, {Min}, {Pinte}, {Thi}, {Kamp},
  {Rab}, {Anthonioz}, {Antonellini}, {Baldovin-Saavedra}, {Carmona}, {Dominik},
  {Dionatos}, {Greaves}, {G{\"u}del}, {Ilee}, {Liebhart}, {M{\'e}nard},
  {Rigon}, {Waters}, {Aresu}, {Meijerink}, \& {Spaans}}]{Woitke2016}
{Woitke}, P., {Min}, M., {Pinte}, C., {et~al.} 2016, \aap, 586, A103,
  \dodoi{10.1051/0004-6361/201526538}

\bibitem[{{Yang} {et~al.}(2013){Yang}, {Ciesla}, \& {Alexander}}]{Yang2013}
{Yang}, L., {Ciesla}, F.~J., \& {Alexander}, C. M.~O.~D. 2013, \icarus, 226,
  256, \dodoi{10.1016/j.icarus.2013.05.027}

\bibitem[{{Yoneda} {et~al.}(2016){Yoneda}, {Tsukamoto}, {Furuya}, \&
  {Aikawa}}]{Yoneda2016}
{Yoneda}, H., {Tsukamoto}, Y., {Furuya}, K., \& {Aikawa}, Y. 2016, \apj, 833,
  105, \dodoi{10.3847/1538-4357/833/1/105}

\bibitem[{{Yurimoto} \& {Kuramoto}(2004)}]{Yurimoto2004}
{Yurimoto}, H., \& {Kuramoto}, K. 2004, Science, 305, 1763,
  \dodoi{10.1126/science.1100989}

\bibitem[{{Zhang} {et~al.}(2017){Zhang}, {Bergin}, {Blake}, {Cleeves}, \&
  {Schwarz}}]{Zhang2017}
{Zhang}, K., {Bergin}, E.~A., {Blake}, G.~A., {Cleeves}, L.~I., \& {Schwarz},
  K.~R. 2017, Nature Astronomy, 1, 0130, \dodoi{10.1038/s41550-017-0130}

\bibitem[{Öberg(2016)}]{Oberg2016}
Öberg, K.~I. 2016, Chemical Reviews, 116, 9631,
  \dodoi{10.1021/acs.chemrev.5b00694}

\end{thebibliography}

\end{document}